\documentclass[journal]{IEEEtran}
\ifCLASSINFOpdf
  \usepackage[pdftex]{graphicx}
  \graphicspath{{../pdf/}{../jpeg/}}
  \DeclareGraphicsExtensions{.pdf,.jpeg,.png}
\else
  \usepackage[dvips]{graphicx}
  \graphicspath{{../eps/}}
  \DeclareGraphicsExtensions{.eps}
\fi

\usepackage{graphicx}
\usepackage{graphics}
\usepackage{epsfig}\usepackage{epstopdf}
\usepackage{epstopdf}
\usepackage{stfloats}
\usepackage[cmex10]{amsmath}
\usepackage{algorithmic}
\usepackage{array}
\usepackage{mdwmath}
\usepackage{mdwtab}
\usepackage{graphicx}
\usepackage{subfigure}
\usepackage{color}
\usepackage{amsfonts,amssymb}
\usepackage[colorlinks,
            linkcolor=red, 
             anchorcolor=red, 
            citecolor=green, 
            ]{hyperref}

\usepackage{algorithm}
\usepackage{algorithmic}

\usepackage{array}
 \usepackage{mathrsfs}

\usepackage{amssymb}
\usepackage{amsmath}
\usepackage{cite}
\usepackage{url}
\usepackage{xcolor}
\usepackage{cite,graphicx,amsmath,amssymb}
\usepackage{subfigure}
\usepackage{fancyhdr}
\usepackage{mdwmath}
\usepackage{mdwtab}
\usepackage{caption}
\usepackage{amsthm}

\newtheorem{theorem}{Theorem}

\newtheorem{lemma}{Lemma}

\newtheorem{corollary}{Corollary}

\begin{document}
\title{Performance Analysis and Optimization of Cooperative Satellite-Aerial-Terrestrial Systems
\thanks{Manuscript received Nov. 15, 2019; revised Mar. 28 and June 18, 2020; accepted June 21, 2020. The associate editor coordinating the review of this paper and approving it for publication was N. Yang. (Corresponding author: Gaofeng Pan.)}
\thanks{G. Pan is with Computer, Electrical and Mathematical Sciences and Engineering Division, King Abdullah University of Science and Technology (KAUST), Thuwal 23955-6900, Saudi Arabia, and he is also with the School of Information and Electronics Engineering, Beijing Institute of Technology, Beijing 100081, China.}
\thanks{J. Ye, Y. Zhang, and M.-S. Alouni are with Computer, Electrical and Mathematical Sciences and Engineering Division, King Abdullah University of Science and Technology (KAUST), Thuwal 23955-6900, Saudi Arabia.}
}

\author{Gaofeng Pan, $Senior Member, IEEE$, Jia Ye, $Student~Member, IEEE$, Yongqiang Zhang, $Student~Member, IEEE$, and Mohamed-Slim Alouini, $Fellow, IEEE$ }


\maketitle

\begin{abstract}
Aerial relays have been regarded as an alternative and promising solution to extend and improve satellite-terrestrial communications, as the probability of line-of-sight transmissions increases compared with adopting terrestrial relays. In this paper, a cooperative satellite-aerial-terrestrial system including a satellite transmitter (S), a group of terrestrial receivers (D), and an aerial relay (R) is considered. Specifically, considering the randomness of S and D and employing stochastic geometry, the coverage probability of R-D links in non-interference and interference scenarios is studied, and the outage performance of S-R link is investigated by deriving an approximated expression for the outage probability. Moreover, an optimization problem in terms of the transmit power and the transmission time over S-R and R-D links is formulated and solved to obtain the optimal end-to-end energy efficiency for the considered system. Finally, some numerical results are provided to validate our proposed analysis models, as well as to study the optimal energy efficiency performance of the considered system.

\end{abstract}



\section{Introduction}
Benefiting from its inherent merits, including large-scale footprint, long communication distance, abundant frequency resource, fast deployment and little dependence on terrestrial facilities, satellite communication has already served as an irreplaceable role in long-distance communications (e.g., television broadcasting), location and navigation, and disaster relief (e.g., weather forecasting). However, direct communication links from the satellite to the ground terminals may not be always available, due to deep fading (e.g., the shadows created by buildings and mountains).

To solve the aforementioned problems, cooperative transmission has been introduced and integrated into satellite networks as an effective strategy to extend the coverage of satellite communications as well as to increase the energy efficiency \cite{Erdelj}. For example, generally, ground stations have been considered to play as relays to aid the communications between the satellite and terrestrial terminals, which can not only raise the coverage level of satellite signals but also provide diversity gain to improve the receiving quality at terrestrial terminals.

So far, there are plenty of researches presented to design and study cooperative satellite-terrestrial communications (CSTC) involving ground relays in terms of bit/symbol error performance \cite{YRuan,AMK,AMK3,LYang,KAn,KAn2,Bhatnagar,Bhatnagar2,Sreng}, outage performance\cite{XYan,Sharma,Cocco}, and physical layer security (PLS) \cite{Bankey,YAi,JDu,BLi,BLi2}.

As an important performance index to quantitatively reflect the event that the transmitted bits/symbols are correctly received at the destination, bit/symbol error probabilities have been investigated for CSTC systems. In integrated satellite and terrestrial networks, the average symbol error rate (ASER) of two transmission modes with co-channel interference under composite multipath/shadowing fading was studied in \cite{YRuan}. The closed-form expression was derived for the SER of $M$-ary phase-shift keying (MPSK) in a two-way amplify-and-forward (AF) satellite system by using a moment-generating function (MGF)-based approach \cite{AMK}. The transmission of orthogonal space-time block codes over a shadowed Rician land mobile satellite link was investigated in \cite{AMK3}, by deriving the expressions of MGF, SER, diversity order, and average capacity. The ASER of MPSK was analyzed for an AF hybrid CSTC network in the presence of co-channel interference \cite{LYang}. The performance of an AF hybrid satellite-terrestrial relay network was investigated by deriving an approximate closed-form expression for ergodic capacity and the analytical lower bound expressions for OP and ASER \cite{KAn}. The ASER of a hybrid satellite-terrestrial decode-and-forward (DF) relay network was evaluated in \cite{KAn2} while considering the effect of co-channel interference. The approximate ASER of MPSK was analyzed for hybrid satellite-terrestrial free space optical AF cooperative link \cite{Bhatnagar}. The ASER was studied for AF relaying hybrid satellite-terrestrial links, while the channel of the terrestrial link between the relay and destination is assumed to suffer Nakagami-$m$ fading \cite{Bhatnagar2}. The authors of \cite{Sreng} evaluated the symbol error probability performance of a hybrid/integrated CSTC network, while MPSK and $M$-ary quadrature amplitude modulation is employed.

Some other researchers paid their attention to the outage performance of CSTC systems to investigate the benefits brought by the relays. In \cite{XYan}, outage probability (OP) and ergodic capacity were studied for downlink hybrid satellite-terrestrial relay networks with a cooperative non-orthogonal multiple access scheme. Closed-form expressions were derived in \cite{Sharma} for the OP of both primary and secondary networks in a hybrid satellite-terrestrial spectrum sharing system. The application of random linear network coding was studied for cooperative coverage extension in land mobile satellite vehicular networks \cite{Cocco}.

It has been proved that, since the information delivery process is extended, the probability that the transmitted information is overheard by eavesdroppers will unavoidably increase when relays are introduced. Then, it is meaningful and necessary to study the impacts of the relays on the information security of CSTC systems. The PLS of a downlink hybrid satellite-terrestrial relay network was in \cite{Bankey}, while \cite{YAi} studied the PLS performance of a hybrid satellite and free-space optical cooperative system. The authors of \cite{JDu} investigated secure communication in a coexistence system of a satellite-terrestrial network and a cellular network through PLS techniques. Ref. \cite{BLi} investigated the secure transmission for cognitive satellite-terrestrial networks where the terrestrial base station serving as a green interference resource is introduced to enhance the security of the satellite link. The authors of \cite{BLi2} investigated the secure transmission in a cognitive satellite-terrestrial network with a multi-antenna eavesdropper, where the interference from terrestrial base stations is introduced to enhance the security of the satellite link.

{\color{black}{Recently, aerial relays, e.g., low-altitude platforms (LAPs) including unmanned aerial vehicles (UAV), balloons and blimps, which operate at varying altitudes in the range of a few dozen meters to a few kilometers, can also assist the communications between the satellite and terrestrial users, leading to cooperative satellite-aerial-terrestrial communication (CSATC) systems. Compared with terrestrial relays, aerial relays are more capable of providing reliable and fast coverage for hazardous scenarios without terrestrial access infrastructure while suffering natural disasters such as floods and earthquakes, and cases of poor communication quality incurred by the lack of line-of-sight (LOS) or deep shadowing, as LOS transmissions can be achieved with high probability by using aerial relays. On the other hand, aerial relays can help some terrestrial terminals (e.g., wireless sensors), which suffer hardware constraints and then cannot directly communicate with the satellite, set up communication links with the satellite. Hence, aerial relays are an alternative choice to realize dependable satellite-terrestrial communications.}} The authors of \cite{PIMRC2016} studied the joint relay selection and power allocation for an orthogonal frequency division multiple access-based hybrid CSATC networks.

Then, one can see that there is almost a blank on the study of CSATC systems. Though researchers have well studied CSTC systems, none of them has considered the randomness of the positions of the satellite and destinations. So, inspired by these observations and similar to our previous work \cite{Zhangyq}, the main purpose of this paper is to investigate the impacts of the randomness of the positions of terrestrial terminals and the satellite to fully understand the performance of CSATC systems, as well as to offer a useful reference to the readers who are interested in studying satellite communication systems.

In this study, a CSATC system consisting of a satellite transmitter (S), a group of terrestrial receivers (D), and a LAP (R) acting as a relay. While employing stochastic geometry theory, the coverage performance of R-D links and the outage performance of S-R link are respectively studied. Then, an optimal problem for transmit power and transmission time allocations is formulated to achieve the optimal end-to-end (e2e) energy efficiency performance. {\color{black}{In a word, logically, the two main tasks of this work is first to understand how system parameters affect the system performance (i.e., coverage and outage performance), and then to optimize the system performance (i.e., energy efficiency).}}

{\color{black}{Technically speaking, compared with existing works on traditional terrestrial cooperative systems, Shadowed-Rician (SR) model presented in \cite{Abdi} is considered in this work, and other than existing works on satellite-terrestrial communication systems, the impact of the randomness of the position of the satellite in three-dimensional space on the performance of the considered system is first studied.}}

The main contributions of this paper are summarized as follows:

{\color{black}{1) The probability density function (PDF) or/and cumulative distribution function (CDF) of the signal-to-noise-ratio (SNR)/signal-to-interference (SIR) over S-R/R-D links are characterized while considering the impacts of both small-scale fading and the randomness of the positions of the terminals;}}

2) The approximated analytical expressions for the coverage probability (CP) of R-D links are derived for non-interference and interference scenarios, respectively;

3) The approximated analytical expression for the OP over S-R link is presented while considering that the satellite is randomly distributed;

{\color{black}{4) The optimal energy efficiency problem is formulated and solved via optimizing the transmit power and time allocation over the two hops to realize the optimal e2e energy efficiency performance of the considered CSATC system.}}

The rest of this paper is organized as follows. In Section II, the considered CSATC system is described. In Section III and IV, the coverage and outage analyses are conducted for R-D and S-R links, respectively. An optimal energy efficiency problem is formulated and solved in Section V. In Section VI, numerical results are presented and discussed. Finally, the paper is concluded with some remarks in Section VII.
\section{System Model}
\begin{figure}[!t]
\centering
\includegraphics[width= 2.8in,angle=0]{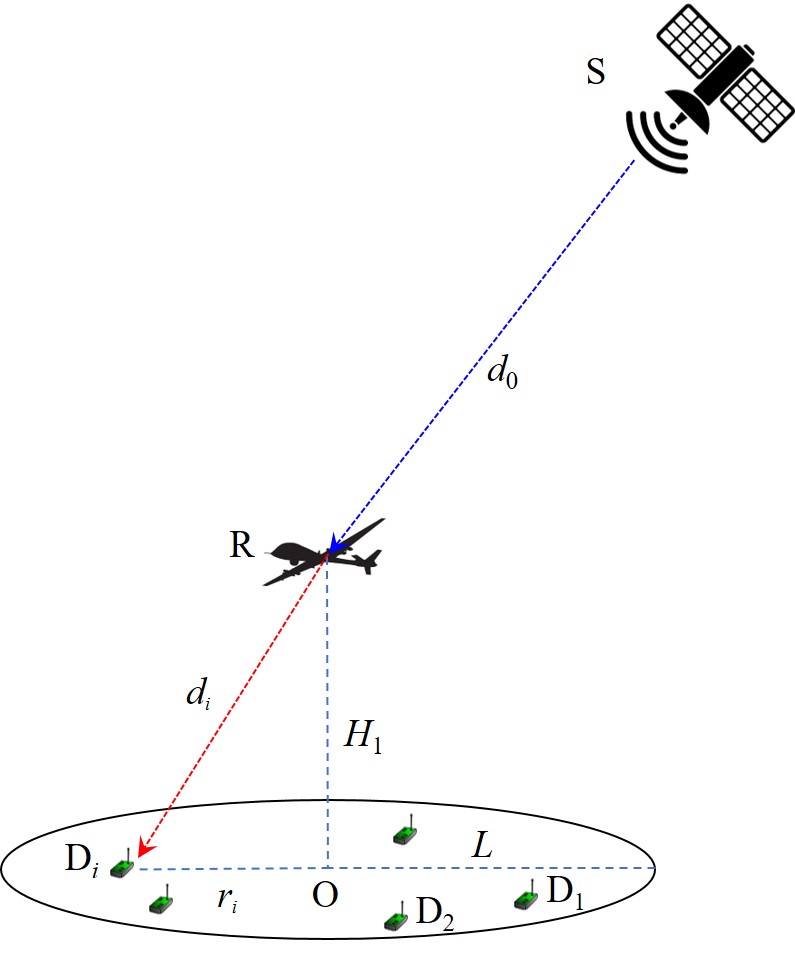}
\caption{System model.}
\label{fig_1}
\end{figure}
In this work, a CSATC system, which consists of a satellite transmitter (S), an aerial relay (R), and a group of terrestrial receivers (${\rm{D}}_i$, $1 \le i \le N$), is considered, as shown in Fig. 1. Specifically, S delivers its information to terrestrial receivers via the help of R, due to the deep fading between S and D, and the hardware limitations of terrestrial receivers (e.g., wireless sensors and common mobile phones) on realizing large-size radio-frequency receiving front end. DF relay scheme is adopted at R to process and forward the received signals from S.

For tractability purposes, in this work we also treat the coverage area of R on the ground as a circle, ${\cal S}$, with radius, $L$, where the projection of R, O, is located at the centre of the circle. The location of terrestrial receivers is modeled as a homogeneous Poisson Point process (PPP) with density $\vartheta$. The number of terrestrial receivers, which denoted by $N$ ($N \le 1$), is Poisson distributed i.e., $P\left\{ {N = k} \right\} = \left( {\mu_{\cal S}^k/k!} \right)\exp \left( { - {\mu_{\cal S}}} \right)$, where ${\mu_{\cal S}} = \pi {L^2}\vartheta $ is the mean measure. Then, the $N$ ($N \le 1$) terrestrial receivers can be modeled as a set of independently and identically distributed points in the circle, ${\cal S}$, denoted by ${\cal W}$. Therefore, the distance between terrestrial receivers and the projection of R (namely, O) can be calculated from ${\cal W}$, the probability density function (PDF) of which can be given by \cite{GPanCL2017,YetWCL}
\begin{align}\label{1}
{f_{\cal W}}\left( w \right) = \frac{\vartheta  }{{\mu \left( {\cal S} \right)}} = \frac{1}{{\pi {L^2}}}.
\end{align}

Furthermore, in this work SR model \cite{Abdi} is adopted to describe the statistical distribution of the satellite link from S to R, which has been proved to be an accurate, practical and applicable tool to evaluate the performance of the satellite propagation environments in various frequency bands, e.g., the UHF-band, L-band, S-band, Ku-band and Ka-band\footnote{In this work, a LAP serves as the aerial relay, the height of which ranges from a few dozen meters to a few kilometers. Then, when the height of the LAP is low, the received signal will suffer both multi-path fading and large-scale fading/shadowing simultaneously; When the height of the LAP is high, the received signal may only experience LOS shadow fading. SR channel model is adopted here to address the channel fading of S-R link, as it is capable of describing the scenarios of pure LOS fading and mixed fading of multi-path fading and shadow fading \cite{Abdi}.}.

Without loss of generality, the PDF of the power gain, ${\left| {{h_{{\rm{SR}}}}} \right|^2}$, over S-R link in SR fading is given as \cite{Abdi}
\begin{align}
{f_{{{\left| {{h_{{\rm{SR}}}}} \right|}^2}}}(x) = \alpha \exp \left( { - \beta x} \right){}_1{F_1}\left( {m;1;\delta x} \right),{\rm{    }}x \ge 0,
\end{align}
where $\alpha ={{{{\left( {\frac{{2bm}}{{2bm + \Omega }}} \right)}^m}} \mathord{\left/
 {\vphantom {{{{\left( {\frac{{2bm}}{{2bm + \Omega }}} \right)}^m}} {\left( {2b} \right)}}} \right.
 \kern-\nulldelimiterspace} {\left( {2b} \right)}}$, $\beta =\frac{1}{{2b}}$, and $\delta =\frac{\Omega }{{2b\left( {2bm + \Omega } \right)}}$, $\Omega$ and $2b$ are the average power of the LOS and multi-path components, respectively, $m$ is the fading severity parameter\footnote{When $m = 0$, it represents the LOS case; when $0 < m < \infty $, it stands for the case of both small-scale fading and LOS; when $ m = \infty $, it denotes the case without LOS.}, and ${}_1{F_1}\left( { \cdot ; \cdot ; \cdot } \right)$ is the confluent hypergeometric function of the first kind.

 Then, it is easy to obtain the received SNR at R as
\begin{align}\label{SNRSR}
{\gamma _{{\rm{SR}}}}&= \frac{{{P_{\rm{S}}}}}{{{\sigma ^2}}}\frac{{{{\left| {{h_{{\rm{SR}}}}} \right|}^2}}}{{d_0^{{n_1}}}}\notag\\
&=\frac{{{\lambda _{\rm{S}}}}}{{d_0^{{n_1}}}},
\end{align}
where ${\lambda _{\rm{S}}}={P_{\rm{S}}}{\left| {{h_{{\rm{SR}}}}} \right|^2}/{\sigma ^2}$, $P_{{\rm{S}}}$ and ${\sigma ^2}$ are the transmit power at S and the average power of the additive white Gaussian noise (AWGN) at R\footnote{In this work, for simplification purposes, it is assumed that the average power of the AWGN at all terminals is same, namely, ${\sigma ^2}$.}, $d_0$ is the distance between S and R, $n_1$ is the path-loss factor over S-R link).

By variable substitution, the PDF and CDF of ${\lambda _{\rm{S}}}$ can be presented as
\begin{align}\label{pdfSR}
{f_{{\lambda _{\rm{S}}}}}(x) = \alpha \sum\limits_{k = 0}^{m - 1} {\frac{{\varsigma \left( k \right)}}{{{\bar \lambda ^{k + 1}}}}{x^k}\exp \left( { - \frac{{\beta  - \delta }}{\bar \lambda }x} \right)}
\end{align}
and
\begin{align}\label{CDFSR}
{F_{{\lambda _{\rm{S}}}}}(x) &= 1 - \alpha \sum\limits_{k = 0}^{m - 1} \frac{{\varsigma \left( k \right)}}{{{\bar \lambda ^{k + 1}}}}\sum\limits_{p = 0}^k {\frac{{k!}}{{p!}}} {{\left( {\frac{{\beta  - \delta }}{\bar \lambda}} \right)}^{ - \left( {k + 1 - p} \right)}}\notag\\
&~~~~\times{x^p}\exp \left( { - \frac{{\beta  - \delta }}{\bar \lambda }x} \right),
\end{align}
respectively, where $\bar \lambda = {P_{{\rm{S}}}}/{\sigma^2}$, $\varsigma \left( k \right) = \frac{{{{\left( { - 1} \right)}^k}{{\left( {1 - m} \right)}_k}{\delta ^k}}}{{{{\left( {k!} \right)}^2}}}$ and ${\left( t \right)_k} = t\left( {t + 1} \right) \cdots \left( {t + k - 1} \right)$ is the Pochhammer symbol \cite{Gradshteyn}.

Similarly, the instantaneous received SNR at the $i$th terrestrial receiver (${\rm{D}}_i$, $1 \le i \le N$ ), can be written as\footnote{In this following parts, we adopt D instead of ${\rm{D}}_i$ in equations by ignoring the subscript for simplification.}

\begin{align}
{\gamma _{{\rm{RD}}}}& = \frac{{{P_{\rm{R}}}}}{{{\sigma ^2}}}\frac{{{{\left| {{h_{{\rm{RD}}}}} \right|}^2}}}{{d_i^{{n_2}}}}\notag\\
& =\frac{{{\lambda _{\rm{R}}}}}{{d_i^{{n_2}}}},
\end{align}
where ${\lambda _{\rm{R}}}={P_{\rm{R}}}{\left| {{h_{{\rm{RD}}}}} \right|^2}/{\sigma ^2}$, $P_{{\rm{R}}}$ is the transmit power at R, $d_i$ is the distance between R and the $i$th terrestrial receiver, $n_2$ is the path-loss factor over R-${\rm{D}}_i$ link, and $h_{\rm{RD}}$ is the channel gain of R-${\rm{D}}_i$ link.

In this work, we assume that the channel among R and the $i$th terrestrial receiver follows Rician distribution. Then, the PDF and CDF of $\lambda_{\rm{R}}$ can be given as \cite{Simon}
\begin{align}\label{PDFSRi}
{f_{{\lambda _{\rm{R}}}}}\left( x \right) &= \frac{{\left( {1 + K} \right)}}{\Omega_{{\rm{R}}} }  \exp \left( { - K - \frac{{x\left( {1 + K} \right)}}{\Omega_{{\rm{R}}} }} \right)\notag\\
&~~~~\times {I_0}\left( {2\sqrt {\frac{{K\left( {1 + K} \right)x}}{\Omega_{{\rm{R}}} }} } \right),x \ge 0
\end{align}
and
\begin{align}\label{CDFSRi}
{F_{{\lambda _{\rm{R}}}}}\left( x \right) = 1 - {Q_1}\left( {\sqrt {2K} ,\sqrt {\frac{{2\left( {1 + K} \right)x}}{\Omega_{{\rm{R}}} }} } \right),
\end{align}
respectively, where $\Omega_{{\rm{R}}}$ is the variance of the signal, $K$ is the Ricci factor, which corresponds to the ratio of the power of the LOS (specular) component to the average power of the scattered component, ${I_0}\left(  \cdot  \right)$ is modified Bessel function  of order 0, ${Q_1}\left( {a,b} \right) = \int\limits_b^\infty  {x  \exp \left( { - \frac{{{x^2} + {a^2}}}{2}} \right)} {I_0}\left( {ax} \right)dx$ is Marcum-$Q$ function.

\section{Coverage Analysis}
In this section, coverage analysis will be carried out under two cases, namely, the ones with/without interference. Also, without loss of generality, we take the $i$th terrestrial receiver, ${\rm{D}}_i$, as the target to study the coverage performance.
\subsection{Non-Interference Scenario}
As one of the most important metrics to evaluate the performance of wireless networks (especially for some application scenarios, e.g., battlefield and hot-spots), CP is defined as the probability that a typical user can achieve some threshold of SNR/SINR. In this section, we will analyze the coverage performance of R-${\rm{D}}_i$ link, while considering the randomness of the positions of terrestrial receivers. Moreover, in this subsection.

In this work, without loss of generality, the $i$th terrestrial receiver, ${\rm{D}}_i$ ($1 \le i \le N$), is considered as the target for the following analysis.

Then, CP can be written as
\begin{align}\label{CPSRi}
{p_c} &= \Pr \left\{ {{\gamma _{{\rm{RD}}}} \ge {\gamma _{{\rm{th}}}}} \right\}\notag\\
& = 1 - \Pr \left\{ {{{\lambda _{\rm{R}}}}   \le {\gamma _{{\rm{th}}}}}{d_i^{{n_2}}} \right\}\notag\\
& = {Q_1}\left( {\sqrt {2K} ,\sqrt {\frac{{2\left( {1 + K} \right){\gamma _{{\rm{th}}}}}}{\Omega_{{\rm{R}}}}d_i^{{n_2}}} } \right),
\end{align}
where $\gamma _{\rm{th}}$ is the predefined threshold.

Using \eqref{1}, we can derive the CDF of ${r_i}$ ($1 \le i \le N$) as
\begin{align}
{F_{{r_i}}}\left( x \right) = \int\limits_0^{2\pi } {\int\limits_0^x {\frac{1}{{\pi {L^2}}}rdrd\theta } }  = \frac{{{x^2}}}{{{L^2}}}.
\end{align}

Therefore, the PDF of  ${r_i}$ ($1 \le i \le N$) can be obtained as ${f_{{r_i}}}\left( x \right) = \frac{{2x}}{{{L^2}}}$, $0 \le x \le L$. Further we can achieve the PDF of ${d_i} = \sqrt {{H_1}^2 + r_i^2} $ ($1 \le i \le N$) as ${f_{{d_i}}}\left( x \right) = \frac{2}{{{L^2}}}x$, ${H_1} \le x \le \sqrt {{H_1}^2 + {L^2}} $.

Considering the randomness of the position of ${\rm{D}}_i$, CP can be further calculated as
\begin{align}\label{PcnonI1}
{p_c} = \int\limits_{{H_1}}^{\sqrt {{H_1}^2 + {L^2}} } {{Q_1}\left( {\sqrt {2K} ,\sqrt {\frac{{2\left( {1 + K} \right){\gamma _{{\rm{th}}}}}}{\Omega_{{\rm{R}}}}{x^{{n_2}}}} } \right)}   {f_{{d_i}}}\left( x \right)dx\notag\\
{\rm{   }} = \frac{2}{{{L^2}}}\int\limits_{{H_1}}^{\sqrt {{H_1}^2 + {L^2}} } {{x}{Q_1}\left( {\sqrt {2K} ,\sqrt {\frac{{2\left( {1 + K} \right){\gamma _{{\rm{th}}}}}}{\Omega_{{\rm{R}}} }{x^{{n_2}}}} } \right)} dx.
\end{align}

As suggested in \cite{Bocus}, Marcum $Q$-function can be approximated as\footnote{{\color{black}{As suggested by Fig. 3 in \cite{Bocus}, the approximation of Marcum $Q$-function given by \eqref{MQapprox} is robust with respect to changes in $a$ from 1 to 5 and can adequately represent the mass of the Marcum $Q$-function while $b$ ranging from 0 to 10.}}}
\begin{align}\label{MQapprox}
{Q_1}\left( {a,b} \right) \approx \exp \left( { - {e^{\upsilon \left( a \right)}}{b^{\mu \left( a \right)}}} \right),
\end{align}
where $\mu \left( a \right) = 2.174 - 0.592a + 0.593{a^2} - 0.092{a^3} + 0.005{a^4}$ and $\upsilon \left( a \right) =  - 0.840 + 0.327a - 0.740{a^2} + 0.083{a^3} - 0.004{a^4}$.

Then, we can obtain
\begin{align}\label{MQapprox2}
{Q_1}\left( {\sqrt {2K} ,\sqrt {\frac{{2\left( {1 + K} \right){\gamma _{{\rm{th}}}}}}{\Omega_{{\rm{R}}}}{x^{{n_2}}}} } \right) \approx \exp \left( { - {e^\upsilon }{\Theta ^{\frac{\mu }{2}}}{x^{\frac{{{n_2}\mu }}{2}}}} \right),
\end{align}
where $\Theta  = \frac{{2\left( {1 + K} \right){\gamma _{{\rm{th}}}}}}{\Omega_{{\rm{R}}}}$, $\mu  = 2.174 - 0.937\sqrt K  + 1.186K - 0.260{K^{\frac{3}{2}}} + 0.010{K^2}$ and $\upsilon  =  - 0.840 + 0.809\sqrt K - 1.480K + 0.235{K^{\frac{3}{2}}} - 0.008{K^2}$.

Using \eqref{MQapprox2} into \eqref{PcnonI1}, the CP for non-interference scenario can be derived as
\begin{align}\label{Pcnon}
{p_c} &\approx  \frac{2}{{{L^2}}}\int\limits_{{H_1}}^{\sqrt {{H_1}^2 + {L^2}} } {{x}\exp \left( { - {e^\upsilon }{\Theta ^{\frac{\mu }{2}}}{x^{\frac{{{n_2}\mu }}{2}}}} \right)} dx\notag\\
&  =  \frac{4}{{L^2{n_2}\mu }}{e^{ - \frac{{4\upsilon }}{{{n_2}\mu }}}}{\Theta ^{ - \frac{2}{{{n_2}}}}}\int\limits_{{y_{\min }}}^{{y_{\max }}} {{y^{\frac{4}{{{n_2}\mu }} - 1}}\exp \left( { - y} \right)} dy\notag\\
& = \frac{4}{{L^2{n_2}\mu }}{e^{ - \frac{{4\upsilon }}{{{n_2}\mu }}}}{\Theta ^{ - \frac{2}{{{n_2}}}}}\left[ \Gamma \left( {\frac{4}{{{n_2}\mu }},{y_{\min }}} \right) \right. \notag\\
&~~~~\left.- \Gamma \left( {\frac{4}{{{n_2}\mu }},{y_{\max }}} \right) \right],
\end{align}
where in the second line we let ${y = {e^\upsilon }{\Theta ^{\frac{\mu }{2}}}{x^{\frac{{{n_2}\mu }}{2}}}}$, ${y_{\max }} = {e^\upsilon }{\Theta ^{\frac{\mu }{2}}}{\left( {{H_1}^2 + {L^2}} \right)^{\frac{{{n_2}\mu }}{4}}}$, ${y_{\min }} = {e^\upsilon }{\Theta ^{\frac{\mu }{2}}}{H_1}^{\frac{{{n_2}\mu }}{2}}$, and $\Gamma \left( {a,x} \right) = \int\limits_x^\infty  {{t^{a - 1}}\exp \left( { - t} \right)dt} $ is the upper incomplete gamma function.

\subsection{Interference Scenario}
In the subsection, we will study the coverage performance of the target system in the presence of a grounded interfering source.

In this work, we assume there is an interfering node existing in the neighbor area of the target terminal, ${\rm{D}}_i$, while R delivering information bits to ${\rm{D}}_i$. This assumption is reasonable and practical, as satellite-terrestrial communications are mainly designed for the scenarios with or even without limited traditional terrestrial communication facilities, e.g., rural areas and battlefields. For example, a macro-cellular typically spans a few kilometers to tens of kilometers to provide radio coverage for rural areas. Then, it is meaningful to assume that there is only one interfering node for our considered CSATC systems.

\begin{figure}[!t]
\centering
\includegraphics[width= 1.8in,angle=0]{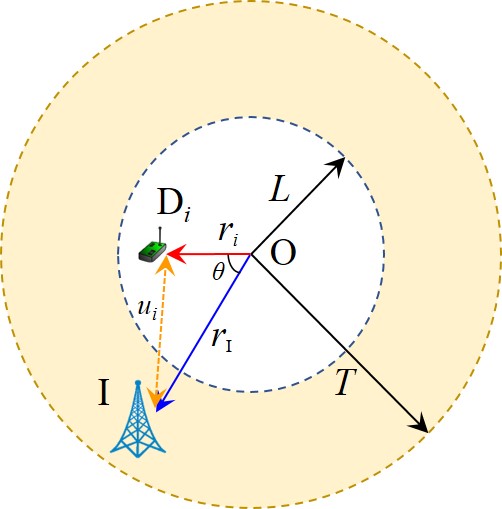}
\caption{Interference scenario.}
\label{fig_2}
\end{figure}
As shown in Fig. 2, it is assumed that the interfering node, I, is uniformly distributed in the circle with original O and radius $T$ ($T > L$). In order to facilitate the following analysis, we denote the polar coordinates of ${\rm{D}}_i$ and I as $(0, r_i)$ and $(\theta, r_I)$, respectively. Then, the distance between ${\rm{D}}_i$ and I can be written as ${u_i} = \sqrt {r_i^2 + r_I^i - 2{r_i}{r_I}\cos \theta } $.

Furthermore, the PDF of $r_I$ and $\theta$ can be easily given as
\begin{align}
{f_{{r_I}}}\left( x \right)=\left\{ {\begin{array}{*{20}{l}}
{\frac{1}{T},}&{{\rm{if~  }}0 \le x \le T;}\\
{0,}&{{\rm{else}}}
\end{array}} \right.
\end{align}
and
\begin{align}
 {f_\theta }\left( x \right)=\frac{1}{{2\pi }},
\end{align}
respectively.

Thus, under this case the received signal at ${\rm{D}}_i$ can be given as
\begin{align}
{y_{{\rm{RD}}}}=\sqrt {\frac{{{P_{\rm{R}}}}}{{d_i^{{n_2}}}}} {h_{{\rm{RD}}}}{x_{{\rm{RD}}}} + \sqrt {\frac{{{P_{\rm{I}}}}}{{u_i^{{n_2}}}}} {h_{{\rm{ID}}}}{x_{\rm{I}}} + z,
\end{align}
where $P_{\rm{I}}$ is the transmit power at I, ${h_{{\rm{ID}}}}$ is the channel gain of the interfering I-${\rm{D}}_i$ link (in this work Rayleigh fading is considered for terrestrial communications, namely, ${f_{{{\left| {{h_{{\rm{ID}}}}} \right|}^2}}}\left( x \right) = {\lambda _{\rm{I}}}\exp \left( { - {\lambda _{\rm{I}}}x} \right)$, $1/{\lambda _{\rm{I}}}$ is the mean of ${\left| {{h_{{\rm{ID}}}}} \right|^2}$), $x_i$ is the transmitted bits from I, $z$ is the complex Gaussian noise at ${\rm{D}}_i$ with average power, ${\sigma ^2}$. As the interfering power is normally much larger than the one of AWGN, in the following we adopt SIR to address and study the effect of interfering signal on the coverage performance of the considered system.

So, the received SIR at ${\rm{D}}_i$ can be derived as
\begin{align}
{\gamma _{{\rm{RD}}}} &= \frac{{\frac{{{P_{\rm{R}}}{{\left| {{h_{{\rm{RD}}}}} \right|}^2}}}{{d_i^{{n_2}}}}}}{{\frac{{{P_{\rm{I}}}{{\left| {{h_{{\rm{ID}}}}} \right|}^2}}}{{u_i^{{n_2}}}}}}\notag\\
&= \frac{{{P_{\rm{R}}}{{\left| {{h_{{\rm{RD}}}}} \right|}^2}}}{{{P_{\rm{I}}}{{\left| {{h_{{\rm{ID}}}}} \right|}^2}}}  \frac{{u_i^{{n_2}}}}{{d_i^{{n_2}}}}\notag\\
& = \frac{{{P_{\rm{R}}}{{\left| {{h_{{\rm{RD}}}}} \right|}^2}}}{{{P_{\rm{I}}}{{\left| {{h_{{\rm{ID}}}}} \right|}^2}}}  \frac{{r_i^2 + r_I^i - 2{r_i}{r_I}\cos \theta }}{{r_i^2 + H_1^2}},
\end{align}
where the path loss factor $n_2 =2$ is only considered in this work for tractability.

\begin{lemma}\label{lem1}
Let $Z = \frac{{{P_{\rm{R}}}{{\left| {{h_{{\rm{RD}}}}} \right|}^2}}}{{{P_{\rm{I}}}{{\left| {{h_{{\rm{ID}}}}} \right|}^2}}}$, the CDF of $Z$ can be written as
\begin{align}
{F_Z}\left( z \right)= {\Gamma _1}{\left( {{\Gamma _2} + \frac{{\Gamma _3}}{z}} \right)^{ - 1}}{}_1{F_1}\left( {1;1;{\Gamma _4}{{\left( {{\Gamma _2} + \frac{{\Gamma _3}}{z}} \right)}^{ - 1}}} \right),
\end{align}
where ${\Gamma _1} = \frac{{1 + K}}{{{\Omega _X}}}\exp \left( { - K} \right)$, ${\Gamma _2} = \frac{{1 + K}}{{{\Omega _X}}}$, ${\Gamma _3} = \frac{{{\lambda _{\rm{I}}}{P_{\rm{R}}}}}{{{P_{\rm{I}}}}}$, and ${\Gamma _4} = \frac{{K\left( {1 + K} \right)}}{{{\Omega _X}}}$.

\begin{proof}
Let $X = {\left| {{h_{{\rm{RD}}}}} \right|^2}$ and $Y = \frac{{{P_{\rm{I}}}}}{{{P_{\rm{R}}}}}{\left| {{h_{{\rm{ID}}}}} \right|^2}$. As ${f_{{{\left| {{h_{{\rm{ID}}}}} \right|}^2}}}\left( x \right) = {\lambda _{\rm{I}}}\exp \left( { - {\lambda _{\rm{I}}}x} \right)$, one can have the CDF of $Y$ as ${F_Y}\left( y \right) = 1 - \exp \left( { - \frac{{{\lambda _{\rm{I}}}{P_{\rm{R}}}}}{{{P_{\rm{I}}}}}y} \right)$.

Similarly, the PDF of $X = {\left| {{h_{{\rm{RD}}}}} \right|^2}$ can be given as
\begin{align}
    {f_X}\left( x \right)& = \frac{{\left( {1 + K} \right)}}{{{\Omega _X}}}  \exp \left( { - K - \frac{{x\left( {1 + K} \right)}}{{{\Omega _X}}}} \right) \notag\\
    &~~~~\times {I_0}\left( {2\sqrt {\frac{{K\left( {1 + K} \right)x}}{{{\Omega _X}}}} } \right),x \ge 0,
\end{align}
where ${\Omega _X} = \Omega {\sigma ^2}/{P_{\rm{R}}}$ is the variance of $X$.

It is easy to write the CDF of $Z$ as
\begin{align}
{F_Z}\left( z \right) &= \Pr \left\{ {Z=\frac{X}{Y} \le z} \right\}\notag\\
&= 1 - \Pr \left\{ {Y \le \frac{X}{z}} \right\}\notag\\
& = \exp \left( { - \frac{{{\lambda _{\rm{I}}}{P_{\rm{R}}}}}{{{P_{\rm{I}}}}}\frac{X}{z}} \right).
\end{align}

Then, using \cite[Eq. (2.15.5.4)]{Prudnikov2}, the CDF of $Z = \frac{X}{Y}$ can be further derived as
\begin{align}
{F_Z}\left( z \right)& = \int\limits_0^\infty  {\exp \left( { - \frac{{{\lambda _{\rm{I}}}{P_{\rm{R}}}}}{{{P_{\rm{I}}}}}\frac{x}{z}} \right)} {f_X}\left( x \right)dx\notag\\
& =  \frac{{\left( {1 + K} \right)}}{{{\Omega _X}}}\int\limits_0^\infty  {\exp \left( {-K - \left( {\frac{{1 + K}}{{{\Omega _X}}} + \frac{1}{z}\frac{{{\lambda _{\rm{I}}}{P_{\rm{R}}}}}{{{P_{\rm{I}}}}}} \right)x} \right)}  \notag \\
&~~~~\times{I_0}\left( {2\sqrt {\frac{{K\left( {1 + K} \right)x}}{{{\Omega _X}}}} } \right)dx\notag\\
&\mathop  = \limits^{t = \sqrt x } 2{\Gamma _1}\int\limits_0^\infty  {{t}}  \exp \left( { - \left( {{\Gamma _2} + \frac{{\Gamma _3}}{z}} \right){t^2}} \right)  {I_0}\left( {2{\Gamma _4^2} t} \right)dt\notag\\
& = {\Gamma _1}{\left( {{\Gamma _2} + \frac{{\Gamma _3}}{z}} \right)^{ - 1}}{}_1{F_1}\left( {1;1;{\Gamma _4}{{\left( {{\Gamma _2} + \frac{{\Gamma _3}}{z}} \right)}^{ - 1}}} \right),
\end{align}
where ${\Gamma _1} = \frac{{1 + K}}{{{\Omega _X}}}\exp \left( { - K} \right)$, ${\Gamma _2} = \frac{{1 + K}}{{{\Omega _X}}}$, ${\Gamma _3} = \frac{{{\lambda _{\rm{I}}}{P_{\rm{R}}}}}{{{P_{\rm{I}}}}}$, and ${\Gamma _4} = \frac{{K\left( {1 + K} \right)}}{{{\Omega _X}}}$.

Then, the proof is completed.
\end{proof}
\end{lemma}

\begin{lemma}
When terminal ${\rm{D}}_i$ is uniformly distributed in the circle with original O and radius $L$, and interfering node I is uniformly distributed in the circle with original O and radius $T$, the CDF of ${\gamma _{{\rm{RD}}}}$ can be derived as
\begin{align}\label{lemma2}
   {F_{{\gamma _{{\rm{RD}}}}}}\left( x \right) &= \frac{{{\Gamma _1}}}{{4\pi}}\sum\limits_g {\sum\limits_j {\sum\limits_k {{\Theta _{g,j,k}}  {}_1{F_1}\left( {1;1;{\Gamma _4}{\Theta _{g,j,k}}} \right)} } } ,
\end{align}
where $\sum\limits_g {\sum\limits_j {\sum\limits_k {} } } $ denotes $\sum\limits_{g = 1}^G {{\vartheta _g}\sum\limits_{j = 1}^H {{\zeta _j}\sqrt {1 - \nu _j^2} } } \left( {{\zeta _j} + 1} \right)$ $\sum\limits_{k = 1}^J {{\iota _k}\sqrt {1 - \kappa _k^2} } \left( {{\kappa _k} + 1} \right)$, ${\varsigma _g} = \cos \left( {\frac{{2g - 1}}{{2G}}\pi } \right)$, ${\vartheta _g} = \frac{\pi }{G}$, ${\eta _1} = \frac{T}{2}$, ${\zeta _j} = \cos \left( {\frac{{2j - 1}}{{2H}}\pi } \right)$, ${\nu _j} = \frac{\pi }{H}$, ${\eta _2} = \frac{L}{2}$, ${\kappa _k} = \cos \left( {\frac{{2k - 1}}{{2H}}\pi } \right)$, ${\iota _k} = \frac{\pi }{J}$, and ${\Theta _{g,j,k}}\left(x  \right) = {\left( {{\Gamma _2} + \frac{{\Gamma _3}}{x}\frac{{{\eta _2}^2{{\left( {{\kappa _k} + 1} \right)}^2} + {\eta _1}^2{{\left( {{\zeta _j} + 1} \right)}^2} - 2{\eta _1}{\eta _2}{\varsigma _g}\left( {{\zeta _j} + 1} \right)\left( {{\kappa _k} + 1} \right)}}{{{\eta _2}^2{{\left( {{\kappa _k} + 1} \right)}^2} + {H_1}^2}}} \right)^{ - 1}}$.
\begin{proof}
Please refer to Appendix I.
\end{proof}
\end{lemma}

\begin{theorem}
Using Lemma 2, the CP for interference scenarios can be presented as
\begin{align}\label{pcinter}
{p_c} &= \Pr \left\{ {{\gamma _{{\rm{RD}}}} \ge {\gamma _{{\rm{th}}}}} \right\}\notag\\
& = 1 - {F_{{\gamma _{{\rm{RD}}}}}}\left( {{\gamma _{{\rm{th}}}}} \right).
\end{align}
\begin{proof}
It is easy to obtain \eqref{pcinter} by using the relationship between CP and the CDF.
\end{proof}
\end{theorem}

\section{Outage analysis for S-R link}
In this section, we will investigate the impact of the randomness of the satellite¡¯s position on the outage performance of the information transmissions over S-R link. It is assumed that S works in the circular orbit with the earth's center, E, which is a sphere with radius, $R_{\rm{E}} = 6371 $ km.

\begin{figure}[!t]
\centering
\includegraphics[width= 2.6in,angle=0]{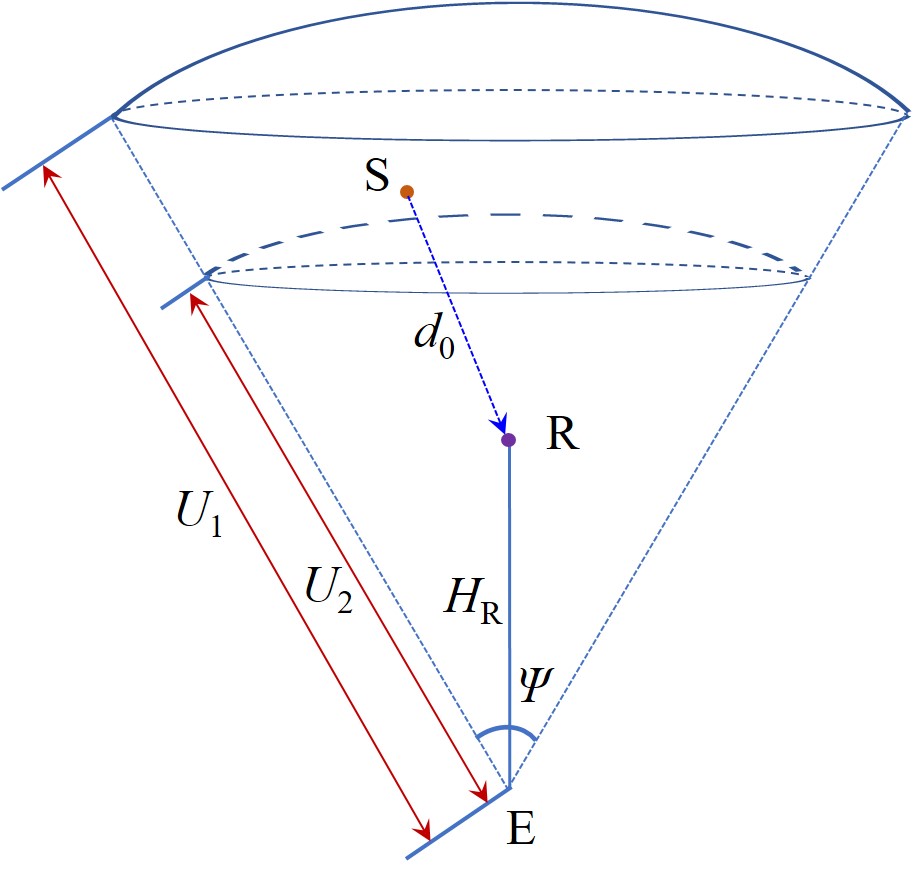}
\caption{Satellite-relay link.}
\label{fig_3}
\end{figure}

As shown in Fig. 3, the satellite, S, is assumed to be uniformly distributed in the space, which is a part that the spherical cone with radius $U_1$ minus the one with radius $U_2$ ($U_1 > U_2$). The two spherical cones are respectively with radius $U_1$ and $U_2$, as well as with sphere central, E, and the same apex angle, $\Psi/2 $\footnote{{\color{black}{$\Psi$ is the apex angle determined by the distribution space of S in which S is able to successfully communicate with R. Hence, we can see that, for practical scenarios, $\Psi$ is decided some system parameters over S-R link, e.g., the transmit power at S, the antenna gains at S and R, the receiver sensitivity at R, the power levels of the noise and interference at R, etc.}}}. The distance between R and E is $H_{\rm{R}}$, which can be written as $H_{\rm{R}} = H_1 + R_{\rm{E}}$. For example, as low earth orbit (LEO) satellites normally revolve at an altitude from the earth¡¯s surface between 160 to 2000 km, then one can have $U_1 = 8371$ km and $U_2 = 6531$ km.

In order to facilitate the following analysis, spherical coordinates are adopted, while E is set as the original. Then, the coordinate of S can be presented as $(r_S, \theta, \psi)$, where $U_2 \le r_S \le U_1$, $0 \le \theta \le \Psi/2$ and $0 \le \psi \le 2\pi$, and the coordinate of R can be written as $(H_{\rm{R}},0,0)$, where $ R_{\rm{E}} \le H_{\rm{R}} \le R_{\rm{E}} + H_{\rm{max, R}}$ and $ H_{\rm{max, R}}$ is the maximum height of R.

Therefore, the distance between S and R, $d_0$, can be presented as
\begin{align}
    {d_0} = \sqrt {r_{\rm{S}}^2 + H_{\rm{R}}^2 - 2{r_{\rm{S}}}{H_{\rm{R}}}\cos \theta }.
\end{align}

So it is obvious that ${U_2} - {H_{\rm{R}}} = {d_{0,\min }} \le {d_0} \le {d_{0,\max }} = \sqrt {U_1^2 + H_{\rm{R}}^2 - 2{U_1}{H_{\rm{R}}}\cos \frac{\Psi }{2}} $. In the following, we consider the case that path-loss factor is 2 to simplify the analysis.

\begin{theorem}
When the position of R is fixed and S is uniformly distributed in the space shown in Fig. 3, the PDF of $d_0^{2}$ can derived as
\begin{align}
{f_{d_0^2}}\left( x \right) &= \frac{3}{{4{H_{\rm{R}}}}} \frac{1}{{1 - \cos \frac{\Psi }{2}}}\frac{1}{{U_1^3 - U_2^3}}\left[ \omega^2 (x) - \rho^2 (x) \right]
\notag\\
&=\tau  \left[ \omega^2 (x) - \rho^2 (x) \right] ,
\end{align}
where $\omega (x) = \min \left\{ {{U_1},{H_{\rm{R}}} + \sqrt x } \right\}$, $\rho (x) = \max \left\{ {{U_2},{H_{\rm{R}}}\cos \frac{\psi }{2} + \sqrt {x - {H_{\rm{R}}}^2{{\sin }^2}\frac{\psi }{2}} } \right\}$,  $ \tau = \frac{3}{{4{H_{\rm{R}}}}}  \frac{1}{{1 - \cos \frac{\Psi }{2}}}\frac{1}{{U_1^3 - U_2^3}}$, and ${x - {H_{\rm{R}}}^2{{\cos }^2}\frac{\Psi }{2} \ge 0}$.

\begin{proof}
Please refer to Appendix II.
\end{proof}
\end{theorem}

 Using \eqref{SNRSR} and \eqref{CDFSR}, the OP for S-R link can be given as
\begin{align}\label{PoutSRdef}
{P_{{\rm{out,SR}}}} &= \Pr \left\{ {{\gamma _{{\rm{SR}}}} \le {\gamma _{\rm{out}}}} \right\}\notag\\
&         = \Pr \left\{ {\frac{{{P_{\rm{S}}}}}{{{\sigma ^2}}}\frac{{{{\left| {{h_{{\rm{SR}}}}} \right|}^2}}}{{d_0^{{2}}}} =\frac{{{\lambda _{\rm{S}}}}}{{d_0^{{2}}}} \le {\gamma _{\rm{out}}}} \right\}\notag\\
&         = 1 - \alpha \sum\limits_{k = 0}^{m - 1} {\frac{{\varsigma \left( k \right)}}{{{\bar\lambda ^{k + 1}}}}\sum\limits_{p = 0}^k {\frac{{k!}}{{p!}}} {{\left( {\frac{{\beta  - \delta }}{\bar\lambda }} \right)}^{ - \left( {k + 1 - p} \right)}}}\notag \\
&~~~~\times { {{\gamma _{\rm{out}}}} ^p}d_0^{{2}p}\exp \left( { - \frac{{\beta  - \delta }}{\bar\lambda }{{\gamma _{\rm{out}}}}d_0^{{2}}} \right),
\end{align}
where ${\gamma _{\rm{out}}}$ is the threshold for the outage events.

Taking the randomness of $d_0^{2}$ in to account and using Theorem 2, the OP for S-R link can be presented as
\begin{align}
{P_{{\rm{out,SR}}}}& = 1 - \alpha \sum\limits_{k = 0}^{m - 1} {\frac{{\varsigma \left( k \right)}}{{{\bar\lambda ^{k + 1}}}}\sum\limits_{p = 0}^k {\frac{{k!}}{{p!}}} {{\left( {\frac{{\beta  - \delta }}{\bar\lambda }} \right)}^{ - \left( {k + 1 - p} \right)}}} \notag\\
&~~~\times{{ {\gamma _{\rm{out}}} }^p}\int\limits_{{d_{0,{{\min }}}^{{2}}}}^{d_{0,\max }^{{2}}} {{x^p}\exp \left( { - \Delta x} \right)} {f_{{d_0^{{2}}}}}\left( x \right)dx\notag\\
&= 1 - \alpha \tau\sum\limits_{k = 0}^{m - 1} {\frac{{\varsigma \left( k \right)}}{{{\bar\lambda ^{k + 1}}}}\sum\limits_{p = 0}^k {\frac{{k!}}{{p!}}} {{\left( {\frac{{\beta  - \delta }}{\bar\lambda }} \right)}^{ - \left( {k + 1 - p} \right)}}} \notag\\
&~~~~ \times{{ {\gamma _{\rm{out}}} }^p} \int\limits_{{d_{0,{{\min }}}^{{2}}}}^{d_{0,\max }^{{2}}} x^p \left[ \omega^2 (x) - \rho^2 (x) \right]\exp \left( { - \Delta x} \right) dx,
\end{align}
where $\Delta  = \frac{{\beta  - \delta }}{\bar \lambda }{\gamma _{\rm{out}}}$.

Again, employing Chebyshev-Gauss quadrature in the first case, the OP for S-R link can be finally written as \eqref{PoutSR} on the top of next page, where ${b_1} = \frac{{d_{0,\max }^{{2}}} - {{d_{0,{{\min }}}^{{2}}}}}{2}$, ${b_2} = \frac{{d_{0,\max }^{{2}}} + {{d_{0,{{\min }}}^{{2}}}}}{2}$, ${\chi _i} = \cos \left( {\frac{{2j - 1}}{{2Q}}\pi } \right)$ and ${\varpi _i} = \frac{\pi }{Q}$.
\begin{figure*}
\begin{align} \label{PoutSR}
{P_{{\rm{out,SR}}}}& = 1 - \alpha \tau b_1\sum\limits_{k = 0}^{m - 1} {\frac{{\varsigma \left( k \right)}}{{{\bar \lambda ^{k + 1}}}}\sum\limits_{p = 0}^k {\frac{{k!}}{{p!}}} {{\left( {\frac{{\beta  - \delta }}{\bar\lambda }} \right)}^{ - \left( {k + 1 - p} \right)}}{{ {\gamma _{\rm{out}}^p} }}} \int\limits_{ - 1}^1 {{{\left( {{b_1}t + {b_2}} \right)}^p}} \notag\\
&~~~~  \times\left[ \omega^2 ({{b_1}t + {b_2}}) - \rho^2 ({{b_1}t + {b_2}}) \right]\exp \left( { - \Delta \left( {{b_1}t + {b_2}} \right)} \right) dt\notag\\
&=1 - \alpha \tau b_1 \sum\limits_{k = 0}^{m - 1} {\frac{{\varsigma \left( k \right)}}{{{\bar\lambda ^{k + 1}}}}\sum\limits_{p = 0}^k {\frac{{k!}}{{p!}}} {{\left( {\frac{{\beta  - \delta }}{\bar\lambda }} \right)}^{ - \left( {k + 1 - p} \right)}}{{{\gamma _{\rm{out}}^p}}}} \sum\limits_{j = 1}^Q {{\varpi _j}\sqrt {1 - \chi _j^2} }{{{{\left( {{b_1}{\chi _j} + {b_2}} \right)}^p}}}\notag\\
&~~~~  \times\left[ \omega^2 ({{b_1}{\chi _j} + {b_2}}) - \rho^2 ({{b_1}{\chi _j} + {b_2}}) \right]\exp \left( { - \Delta \left( {{b_1}{\chi _j} + {b_2}} \right)}\right)
\end{align}
\rule{18cm}{0.01cm}
\end{figure*}

\begin{corollary}
When DF scheme is adopted at R, the e2e OP for S-R-${\rm{D}}_i$ link can be finally obtained as
\begin{align}
    P_{e2e} = 1 - p_c(1 - {P_{{\rm{out,SR}}}}),
\end{align}
where $p_c$ is presented as \eqref{Pcnon} and \eqref{pcinter} for non-interference and interference scenarios, respectively, and ${P_{{\rm{out,SR}}}}$ is given as \eqref{PoutSR}.
\end{corollary}

\section{The e2e Energy Efficiency Optimization}
It is well-known that satellites and aerial relays are resource-constrained systems, especially suffering rigid power budget to realize the designed operating life. Though satellites are normally equipped with solar panels to charge the batteries, the energy renewing process is restricted by the satellites' orbiting movements. Therefore, it is vital for the considered CSATC systems to efficiently and effectively exploit the limited power resource at both satellites and aerial relays.

As presented in Fig. 1, intuitively, there exists an optimal transmit power and transmission time allocation over S-R and R-D hops to achieve optimal e2e energy efficiency for the considered CSATC system. Because the transmit power and transmission time over S-R and R-${\rm{D}}_i$ hops are directly related to the total energy and time consumption to realize the delivery of the same amount of information bits over both S-R and R-${\rm{D}}_i$ hops. Then, in this section, we will present an optimization problem in terms of the transmit power and transmission time over S-R and R-${\rm{D}}_i$ hops to realize the optimal e2e energy efficiency performance of S-R-${\rm{D}}_i$ link.

Furthermore, in this study, we only concern the energy consumed on transmitting information bits and ignore that spent on processing the signals (e.g., encoding and decoding), as radio frequency communications have been proved as the main parts of energy consumption at the terminals.

\subsection{Problem Formation}
Similar to \cite{PanTcom2017}, in this work energy efficiency is adopted to evaluate the efficiency of the energy consumption on delivering the information bits, which is defined as the number of delivered bits over S-R-${\rm{D}}_i$ link with unit-joule consumption. Therefore, the following optimization problem is considered: How to optimally allocate the transmit power and transmission time over S-R and R-${\rm{D}}_i$ links so that the energy efficiency of the considered system is maximized.

\begin{equation}\label{Equ.1}
\hspace{-0.2in}\mathbf{OPT-1:}\hspace{0.3in}{\max\limits_{\mathcal{P},\mathcal{T}}}\hspace{0.1in}\eta =\frac{D_{{\rm{SD}}}}{P_{{\rm{S}}}T_{{\rm{S}}}+ P_{{\rm{R}}}T_{{\rm{R}}}}
\end{equation}
Subject to
\begin{eqnarray*}
C1 &:&T_{{\rm{S}}}B_{{\rm{S}}}\log_2(1+P_{{\rm{S}}}\gamma_{{\rm{SR}}}) = D_{{\rm{SD}}},\\
C2 &:&T_{{\rm{R}}}B_{{\rm{R}}}\log_2(1+P_{{\rm{R}}}\gamma_{{\rm{RD}}}) = D_{{\rm{SD}}}, \\
C3 &:&T_{{\rm{S}}} + T_{{\rm{R}}} \leq T , \\
C4 &:&T_{{\rm{S}}}\geq \frac{D_{{\rm{SD}}}}{B_{{\rm{S}}}\log_2(1+P_{{\rm{S}}}^{\max}\gamma_{{\rm{SR}}})}, \\
C5 &:&T_{{\rm{R}}}\geq \frac{D_{{\rm{SD}}}}{B_{{\rm{R}}}\log_2(1+P_{{\rm{R}}}^{\max}\gamma_{{\rm{RD}}})},\end{eqnarray*}%
where $\mathcal{P} = \{P_{{\rm{S}}},P_{{\rm{R}}}\}$ is the power allocation policy, and  $\mathcal{T} = \{T_{{\rm{S}}},T_{{\rm{R}}}\}$ is the time allocation policy. $P_{{\rm{S}}}^{\max}$ and $P_{{\rm{R}}}^{\max}$ denotes the maximum transmit power at S and R, respectively. $\gamma_{{\rm{SR}}} = \frac{|h_{{\rm{SR}}}|^2}{\sigma^2d_0^{n_1}}$, $\gamma_{{\rm{RD}}} = \frac{|h_{{\rm{RD}}}|^2}{\sigma^2d_i^{n_2}}$ for non-interference case, and $\gamma_{{\rm{RD}}} =\frac{{\left| {{h_{{\rm{RD}}}}} \right|}^2}{{{P_{\rm{I}}}{{\left| {{h_{{\rm{ID}}}}} \right|}^2}}} \frac{{u_i^{{n_2}}}}{{d_i^{{n_2}}}}$ for interference case. C1 and C2 are the transmission rate constraints, which guarantee the amount of the transmitted data over the two hops is equal to the requirement, $D_{{\rm{SD}}}$. C3 restricts that the total transmission time for the two hops is bounded by the maximum time $T$ for the e2e transmission. C4 and C5 ensures the low bounds of the transmission time for the two hops, respectively. The coupling between $\mathcal{P}$ with $\mathcal{T}$ results in an obstacle to address the problem $\mathbf{OPT-1}$. However, by exploiting Lemma \ref{lemma3} in the following, we can derive the optimal solution by solving the convex optimization problem $\mathbf{OPT-2}$ as follows

\begin{align}\label{Equ.2}
&\mathbf{OPT-2:}\notag\\
&~~~~~~{\min\limits_{\mathcal{T}}}~\eta_t = {\frac{T_{{\rm{S}}}}{\gamma_{{\rm{SR}}} }g\left(\frac{D_{{\rm{SD}}}}{T_{{\rm{S}}}},B_{{\rm{S}}}\right)+ \frac{T_{{\rm{R}}}}{\gamma_{{\rm{RD}}} }g\left(\frac{D_{{\rm{SD}}}}{T_{{\rm{R}}}},B_{{\rm{R}}}\right)}
\end{align}
Subject to C3, C4, and C5.

\begin{lemma}\label{lemma3}
 We can derive the solution of problem $\mathbf{OPT-1}$ by solving the convex optimization problem $\mathbf{OPT-2}$.
\begin{proof}
We first define a function $g(x,B) = 2^{\frac{x}{B}}-1$. Then, $P_{{\rm{S}}} = \frac{1}{\gamma_{{\rm{SR}}} }g\left(\frac{D_{{\rm{SD}}}}{T_{{\rm{S}}}},B_{{\rm{S}}}\right)$ and $P_{{\rm{R}}} = \frac{1}{\gamma_{{\rm{RD}}} }g\left(\frac{D_{{\rm{SD}}}}{T_{{\rm{R}}}},B_{{\rm{R}}}\right)$.

Further, we can rewrite the objective function of $\mathbf{OPT-1}$ as
\begin{equation}\label{Equ.3}
\hspace{0.3in}{\max\limits_{\mathcal{P},\mathcal{T}}}\hspace{0.2in}\eta =\frac{D_{{\rm{SD}}}}{\frac{T_{{\rm{S}}}}{\gamma_{{\rm{SR}}} }g\left(\frac{D_{{\rm{SD}}}}{T_{{\rm{S}}}},B_{{\rm{S}}}\right)+ \frac{T_{{\rm{R}}}}{\gamma_{{\rm{RD}}} }g\left(\frac{D_{{\rm{SD}}}}{T_{{\rm{R}}}},B_{{\rm{R}}}\right)}.
\end{equation}

It is obvious that the maximum energy efficiency $\eta^*$ is equal to $\frac{D_{{\rm{SD}}}}{\eta_t^*}$.

Since the second derivative of $g(x,B)$, i.e., $\frac{\mathrm{d}^2g(x,B)}{\mathrm{d}x}$, is always non-negative, $g(x,B)$ is a convex function w.r.t $x$. The perspective function of $g(x,B)$ i.e., $xg\left(\frac{1}{x},B\right)$ is also a convex function\cite{Boyd}. Thus, the objective function of $\mathbf{OPT-2}$ is the sum of two convex functions, and the constraints are linear. As a result, the transformed problem $\mathbf{OPT-2}$ is a convex optimization problem w.r.t $\mathcal{T}$.
\end{proof}
\end{lemma}

\subsection{Solution of the Problem OPT-2}
The Lagrangian function of the problem $\mathbf{OPT-2}$ is given by
\begin{align}\label{Equ.4}
     L(\lambda,T_{{\rm{S}}},T_{{\rm{R}}}) = & {\frac{T_{{\rm{S}}}}{\gamma_{{\rm{SR}}} }g\left(\frac{D_{{\rm{SD}}}}{T_{{\rm{S}}}},B_{{\rm{S}}}\right)+ \frac{T_{{\rm{R}}}}{\gamma_{{\rm{RD}}} }g\left(\frac{D_{{\rm{SD}}}}{T_{{\rm{R}}}},B_{{\rm{R}}}\right)}\notag\\
     & + \lambda(T_{{\rm{S}}} + T_{{\rm{R}}}- T),
\end{align}
where $\lambda$ is the Lagrangian multiplier corresponds to the total time constraint.

 Then, the dual problem of \eqref{Equ.4} can be presented as
\begin{equation}\label{Equ.5}
    {\max\limits_{\lambda\geq0}}\hspace{0.1in}{\min\limits_{\mathcal{T}}}\hspace{0.1in}L(\lambda,T_{{\rm{S}}},T_{{\rm{R}}}) .
\end{equation}

By solving the dual problem \eqref{Equ.5}, we can derive the solution of problem \eqref{Equ.2}. To this end, we first solve the inner problem for fixed $\lambda$ and then update the Lagrangian multiplier $\lambda$ by subgradient method until the Lagrangian multiplier converges.

Using KKT conditions, the time allocation policy for the first phase transmission, i.e., from S to R, is given by
\begin{equation}\label{Equ.6}
    T_{{\rm{S}}}^* = \Big[\hat{T_{{\rm{S}}}}\Big]_{{T_{{\rm{S}}}}_{\min}},
\end{equation}
where $[x]_a$ is an operator which is defined as $\max(a,x)$, ${T_{{\rm{S}}}}_{\min} = \frac{D_{{\rm{SD}}}}{B_{{\rm{S}}}\log2(1+P_{{\rm{S}}}^{\max}\gamma_{{\rm{SR}}})}$,  and $\hat{T_{{\rm{S}}}}$ denotes the solution of the following equation
\begin{equation}\label{Equ.7}
        1-\frac{\ln2D_{{\rm{SD}}}}{T_{{\rm{S}}}B_{{\rm{S}}}}= (1-\gamma_{{\rm{SR}}}\lambda) \left(\frac{1}{2}\right)^{\frac{D_{{\rm{SD}}}}{B_{{\rm{S}}}T_{{\rm{S}}}}}.
\end{equation}

Similarly, the time allocation policy for the second phase transmission, i.e., from S to ${\rm{D}}_i$, is given by
\begin{align}\label{Equ.8}
    T_{{\rm{R}}}^* = \Big[\hat{T_{{\rm{R}}}}\Big]_{{T_{{\rm{R}}}}_{\min}},
\end{align}
where ${T_{{\rm{R}}}}_{\min} = \frac{D_{{\rm{SD}}}}{B_{{\rm{R}}} \log2(1+P_{{\rm{R}}}^{\max}\gamma_{{\rm{RD}}})}$, and $\hat{T_{{\rm{R}}}}$ denotes the solution of the following equation
\begin{equation}\label{Equ.9}
       1-\frac{\ln2D_{{\rm{SD}}}}{T_{{\rm{R}}}B_{{\rm{R}}}}= (1-\gamma_{{\rm{RD}}}\lambda)\left(\frac{1}{2}\right)^{\frac{D_{{\rm{SD}}}}{B_{{\rm{R}}}T_{{\rm{R}}}}}  .
\end{equation}

Since \eqref{Equ.7} and \eqref{Equ.9} are transcendental equations, which do not have a closed-form solution generally, we can use some numerical methods (e.g., Newton method) to find the approximate solution. Moreover, there are some packets on various platforms to solve such a simple form transcendental equation, such as Mathematica.

As for the high-level problem of \eqref{Equ.5}, we update the Lagrangian multiplier via the subgradient method
\begin{equation}\label{Equ.10}
    \lambda(j+1) = \Big[\lambda(j) + \varphi(j)\big( T_{{\rm{S}}} + T_{{\rm{R}}} - T \big) \Big]_0,
\end{equation}
where $j\geq 0$ denotes the iteration index, and $\varphi(j)$ represents the positive diminishing step size.

The pseudocode of the proposed algorithm is shown in Algorithm \ref{alg:1}. The proposed iterative algorithm consists of only one loop, it has a polynomial time complexity, i.e., $O(N)$.

\begin{algorithm}[H]
\caption{Iterative Algorithm for \textbf{OPT-2} Problem}
\label{alg:1}
 \begin{algorithmic}[1]
 \STATE {\bf Initialization:}Initialize $\lambda(0)$, and sets $j=0$\\
 \REPEAT
 \STATE  Calculate $T_{{\rm{S}}}^*$ and $T_{{\rm{R}}}^*$ based on \eqref{Equ.6} and \eqref{Equ.8}  \\
 \STATE Update $\lambda(j)$ by using \eqref{Equ.10} ; \\
 \STATE $j = j+1$
 \UNTIL{Sequence $\lambda$ converges }\\
 \STATE  Calculate the $\eta_t^*$ by \eqref{Equ.2}
 \RETURN $T_{{\rm{S}}}^*$, $T_{{\rm{R}}}^*$ and $\eta^* = \frac{D_{{\rm{SD}}}}{\eta_t^*}$
  \end{algorithmic}
\end{algorithm}

\begin{figure}[!t]
\centering
\includegraphics[width= 3.5in,angle=0]{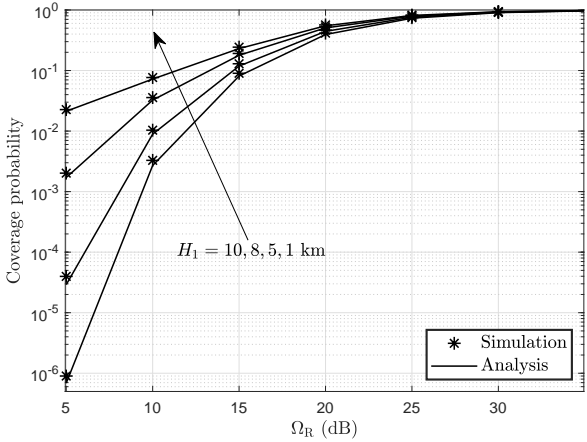}
\caption{CP over R-D link for various $H_1$.}
\label{fig_4}
\end{figure}

\begin{figure}[!t]
\centering
\includegraphics[width= 3.5in,angle=0]{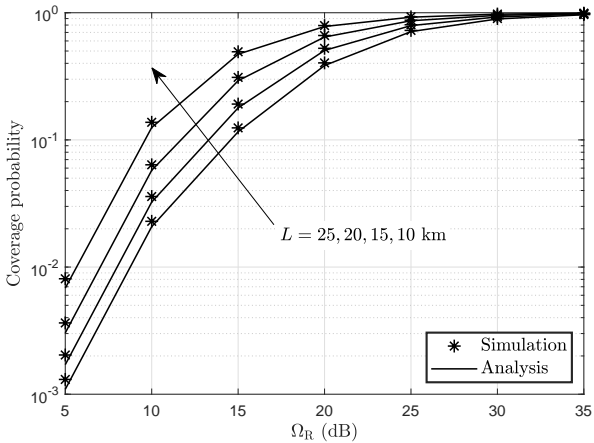}
\caption{CP over R-D link for various $L$.}
\label{fig_5}
\end{figure}

\section{Numerical Results}
In this section, numerical results will be provided to study the performance of the considered system, as well as to verify the proposed analytical models and optimal designs. In the following, we use D instead of ${\rm{D}}_i$ for simplification, and path-loss factors for S-R and R-D links are set as 2. Furthermore, we run $1 \times 10^6$ times of the realizations of the considered system and $1 \times  10^6$ trials of Monte-Carlo simulations, to model the randomness of the positions of the considered terminals and channel gains over each link.

\subsection{Coverage Performance Over S-R Link}
In this subsection, $\gamma_{\rm{th}} = 0$ dB, and two scenarios, non-interference and interference scenarios, will be studied for various the average channel power gain of R-D link ($\Omega_{{\rm{R}}}$), respectively.
\subsubsection{Non-interference scenarios}
Main parameters adopted in this case are set as: $P_{{\rm{R}}} = 1$ dB, $K = -10$ dB, $L = 20$ km, and $H_1 = 5$ km.

In Figs. 4 and 5, the CP over R-D links are presented for various of $H_1$ and $L$, respectively. One can see that in both figures CP can be improved while $\Omega_{{\rm{R}}}$ increasing. Because a large $\Omega_{{\rm{R}}}$ presents a large average power of the received signal, which means that the received SNR at D increases.

Also, it is obvious that a small $H_1$/$L$ leads to a large CP. This observation comes from the fact that a small $H_1$ denotes a low path-loss for the signal transmitted over R-D link, and a small $L$ shows a small distributed area for D, leading to small path-loss over R-D link.

\subsubsection{Interference Scenarios}
Main parameters adopted in this case are set as: $G=H=J=50$, $P_{{\rm{R}}} = 1$ dB, $K = -10$ dB, $L = 5$ km, $H_1 = 5$ km, and $T = 30$ km.
\begin{figure}[!t]
\centering
\includegraphics[width= 3.5in,angle=0]{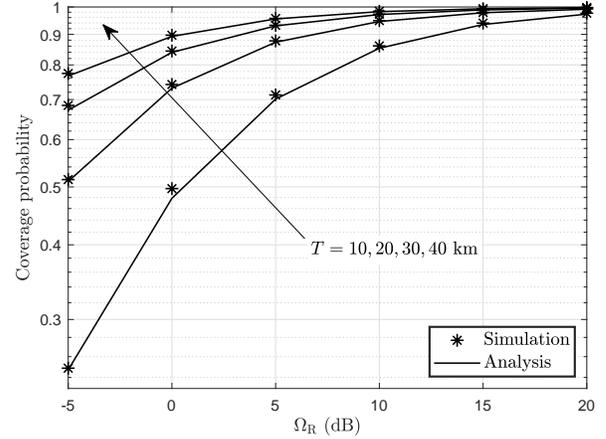}
\caption{CP over R-D link for various $T$.}
\label{fig_6}
\end{figure}

Fig. 6 shows the impact of the size of the distribution area for the interfering node on the coverage performance of the considered system. we can observe that the CP with a large $T$ outperforms the one with a small $T$. It can be explained that a large $T$ represents a large distribution area for the interfering node, which exhibits a large probability of large path-loss over the interfering link.

\begin{figure}[!t]
\centering
\includegraphics[width= 3.5in,angle=0]{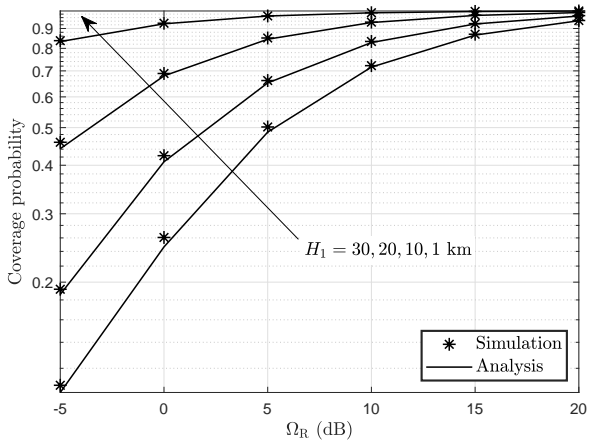}
\caption{CP over R-D link for various $H_1$.}
\label{fig_7}
\end{figure}

In Fig. 7, the influence of $H_1$ on the coverage performance of the considered system is presented for the interference scenario. It is easy to find out that $H_1$ shows a negative effect on the coverage performance of the considered system, as the CP with a low $H_1$ outperforms the one with a large $H_1$. This finding can also be explained by the reason proposed for the observations in Fig. 4.

Furthermore, one can clearly find that simulation and numerical results agree with each other very well, which verifies the accuracy of the proposed analytical model presented in Section III.

\subsection{Outage Performance}
In this subsection, we set $\gamma_{\rm{out}} =0$ dB, $\Psi = \pi/3$, $m =2$, $R_{\rm{E}} = 6371 $ km, $U_1 = 8371$ km and $U_2 = 6531$ km, and the outage performance over S-R link and the e2e outage performance of the considered system will be investigated for various the average power of the LOS components over S-R link ($\Omega$) and various the average channel power gain of R-D link ($\Omega_{{\rm{R}}}$), respectively.
\subsubsection{OP over S-R link}
Main parameters adopted here are set as: $P_{{\rm{S}}} = 30$ dB, $b = 10$ dB, $P = 50$.

\begin{figure}[!t]
\centering
\includegraphics[width= 3.5in,angle=0]{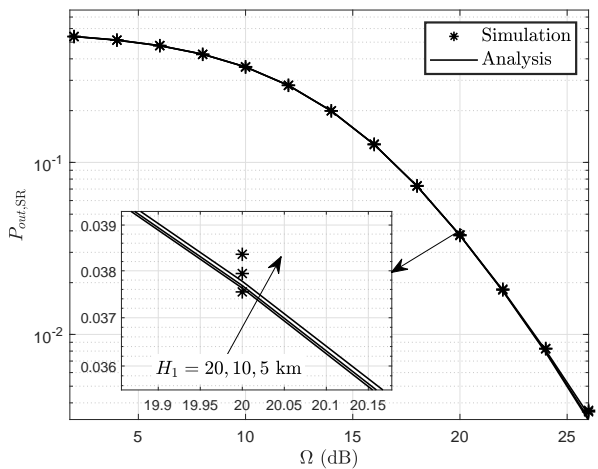}
\caption{OP over S-R link for various $H_1$.}
\label{fig_8}
\end{figure}

In Fig. 8, the outage performance over S-R link is studied for various $H_1$. It can be seen that the height of R, $H_1$, exhibits quite a weak impact on the outage performance over S-R link since the differences among the OP for various $H_1$ from 5 km to 20 km is very narrow (roughly on the orders of $10^{-3}$, as suggested by Fig. 8). Since the distance of S-R link is on the order of hundreds to thousands of kilometers, the variations of the height of R, $H_1$, from tens to thousands of meters cannot produce a significant effect on the outage performance over S-R link.

\begin{figure}[!t]
\centering
\includegraphics[width= 3.5in,angle=0]{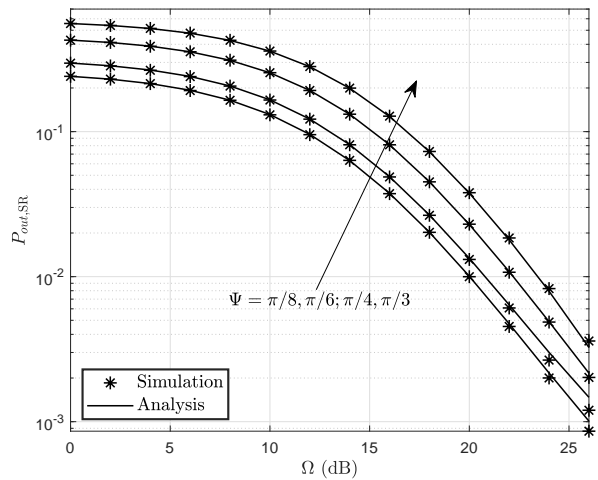}
\caption{OP over S-R link for various $\Psi$.}
\label{fig_9}
\end{figure}

\begin{figure}[!t]
\centering
\includegraphics[width= 3.5in,angle=0]{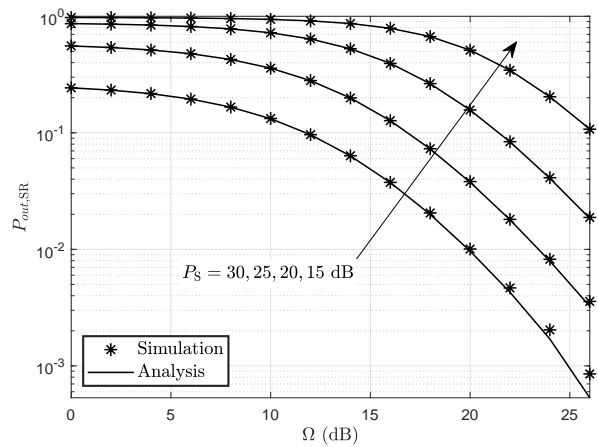}
\caption{OP over S-R link for various $P_{{\rm{S}}}$.}
\label{fig_10}
\end{figure}

Fig. 9 presents the outage performance over S-R link for various $\Psi$, while $\Omega$ increasing. Obviously, $\Psi$ shows a negative impact on the outage performance over S-R link, as a large $\Psi$ means a large distribution space for S, resulting in large path-loss for S-R link.

The influence of the transmit power at S, $P_{{\rm{S}}}$, on the outage performance over S-R link is depicted in Fig. 10. As expected, $P_{{\rm{S}}}$ exhibits a positive effect, since increasing $P_{{\rm{S}}}$ can definitely improve the received SNR at R with no doubts.

Moreover, as presented in Figs. 8-10, the OP over S-R link is improved, while $\Omega$ increasing. Because $\Omega$ represents the average power of the LOS components of the received signal at R.

Finally, we can also easily see that simulation and numerical results match well with each other, which confirms the correctness of the proposed analytical model presented in Section IV.

\begin{figure}[!t]
\centering
\includegraphics[width= 3.5in,angle=0]{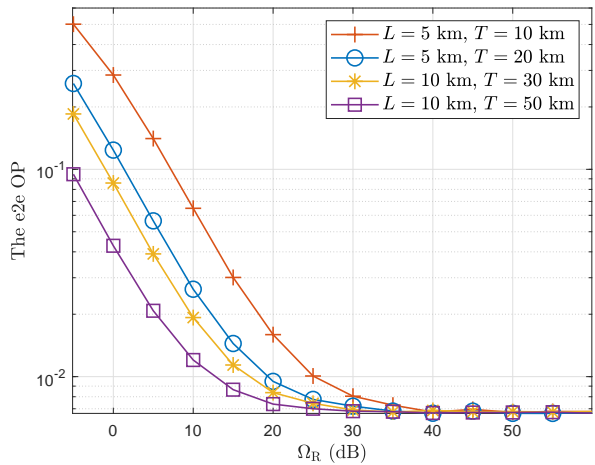}
\caption{The e2e OP for interference scenarios with various combinations of $L$ and $T$.}
\label{fig_11}
\end{figure}

\begin{figure}[!t]
\centering
\includegraphics[width= 3.5in,angle=0]{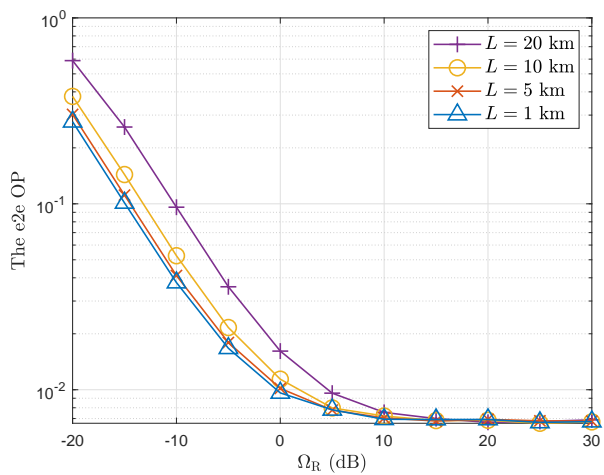}
\caption{The e2e OP for non-interference scenarios with various $L$.}
\label{fig_12}
\end{figure}

\subsubsection{The e2e OP}
In this part, we will present some simulation results of the e2e outage performance of the considered system shown in Fig. 1. Main parameters adopted in this part are set as: $H_1$ = 10 km, $L = 5$ km, $T$ = 50 km, $P_I = P_{{\rm{R}}} = 1$ dB, $P_{{\rm{S}}} = 30$ dB, $K = -10$ dB.

As depicted in Figs. 11 and 12, the size of the distribution area of D, $L$, shows a negative influence on the e2e OP of the considered system in both interference and non-interference scenarios, since a small $L$ means a large probability of small path-loss over R-D link.

In Fig. 11, we can see that increasing the range of the distribution area of the interfering node can improve the e2e OP of the considered system, as the probability that the interfering distance from the interfering node to D increases will enlarge, leading to the decreased interfering power and further the improved SIR at D.

\subsection{Optimal Energy Efficiency Design}

\begin{figure}[!t]
\centering
\includegraphics[width= 3.5in,angle=0]{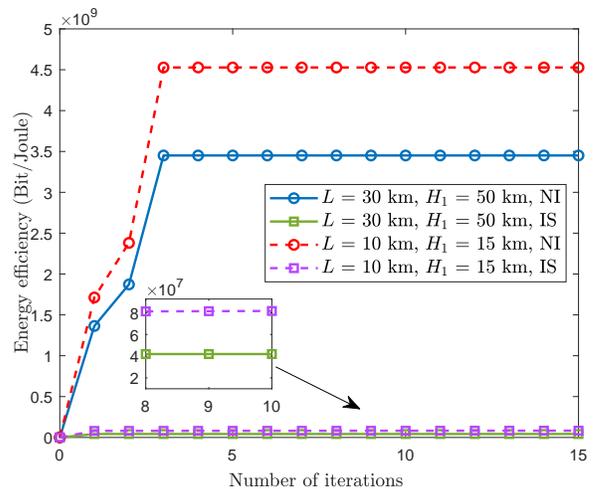}
\caption{The optimal energy efficiency versus the number of the iterations for Algorithm \ref{alg:1}.}
\label{fig_13}
\end{figure}

\begin{figure}[!t]
\centering
\includegraphics[width= 3.5in,angle=0]{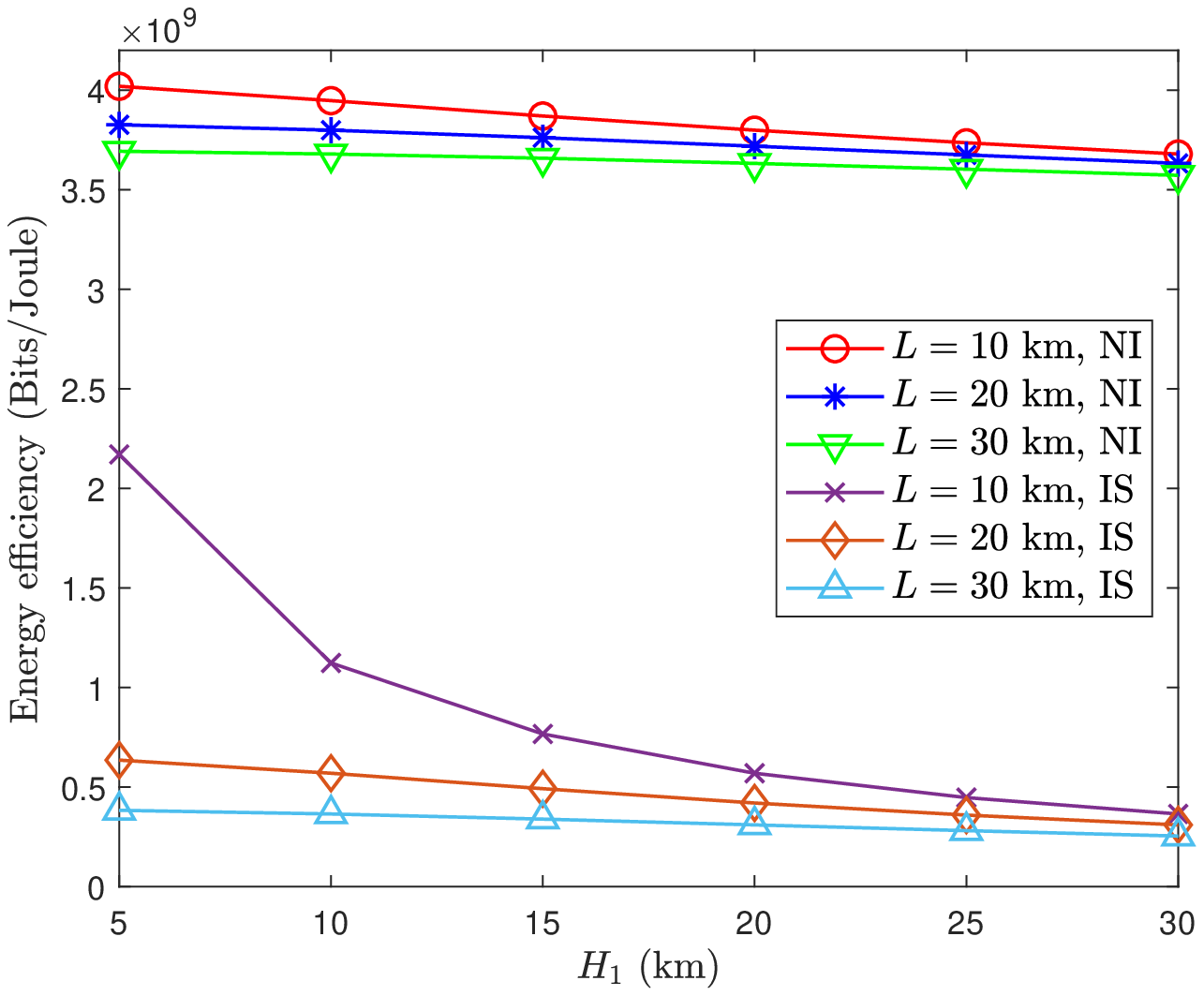}
\caption{The optimal energy efficiency for various $L$.}
\label{fig_14}
\end{figure}

\begin{figure}[!t]
\centering
\includegraphics[width= 3.5in,angle=0]{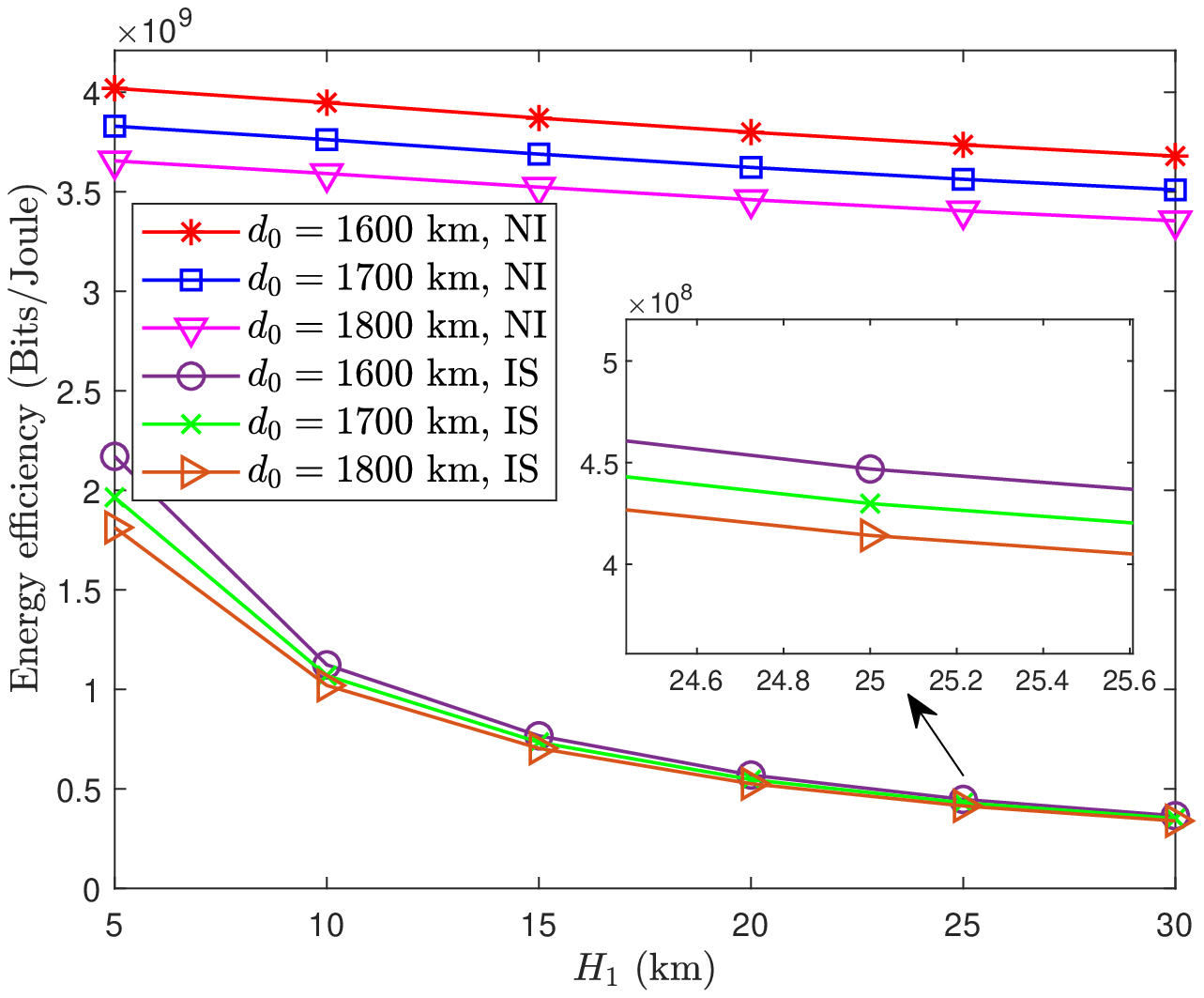}
\caption{The optimal energy efficiency for various $d_0$.}
\label{fig_15}
\end{figure}

\begin{figure}[!t]
\centering
\includegraphics[width= 3.5in,angle=0]{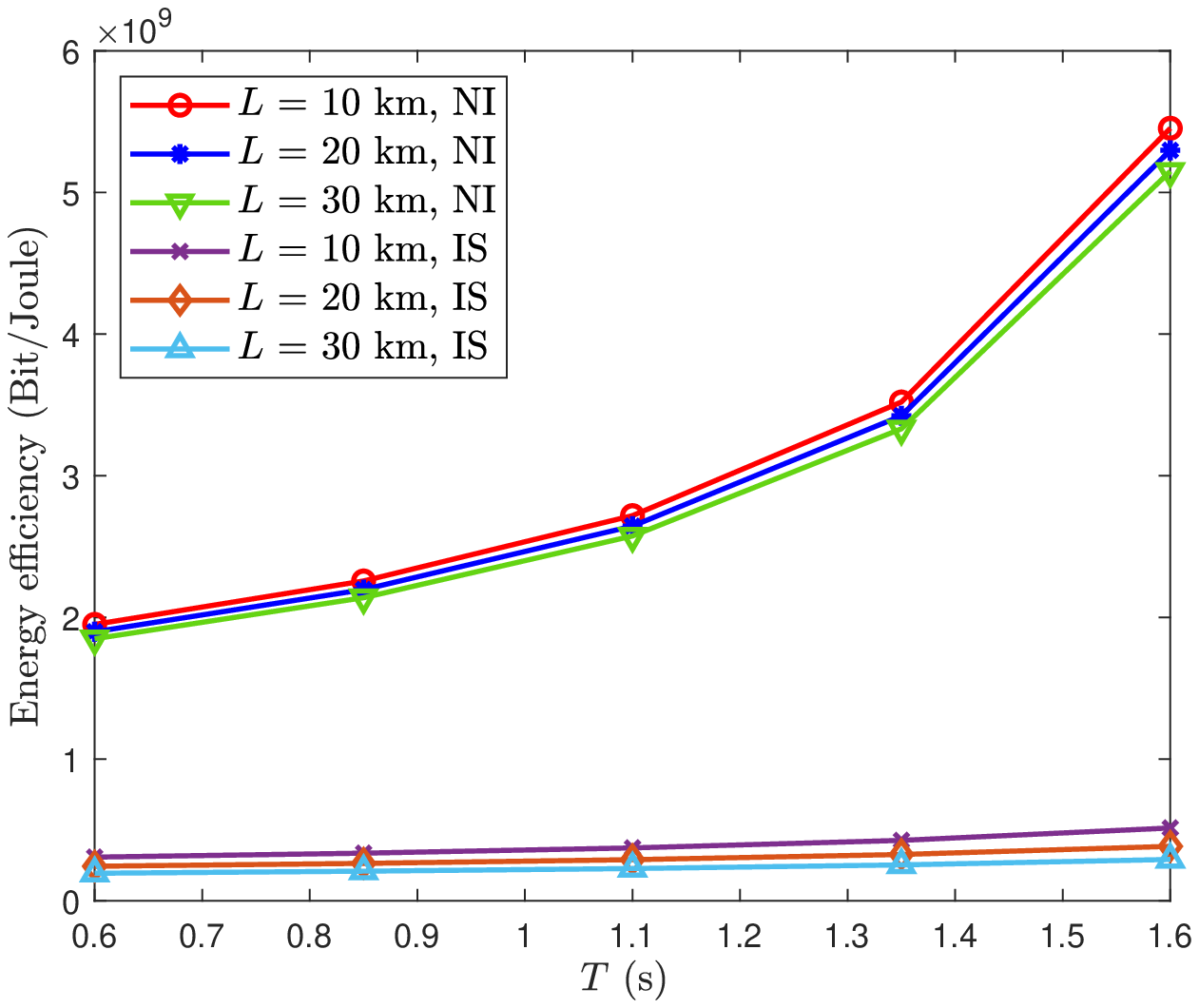}
\caption{The optimal energy efficiency for various $T$.}
\label{fig_16}
\end{figure}

\begin{figure}[!t]
\centering
\includegraphics[width= 3.5in,angle=0]{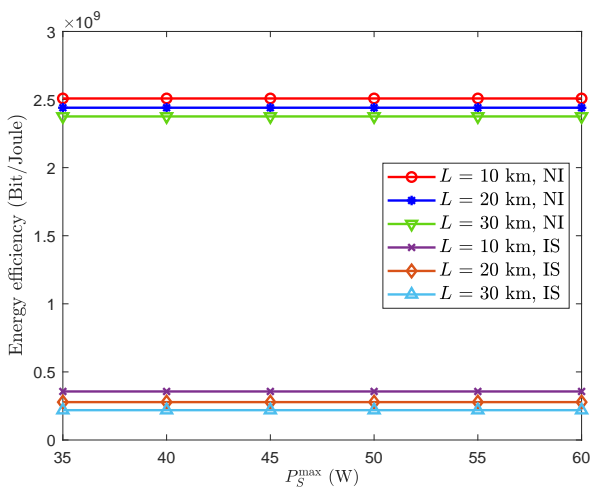}
\caption{The optimal energy efficiency for various $P_s^{\max}$.}
\label{fig_17}
\end{figure}

In this subsection, we will present the e2e energy efficiency for non-interference and interference scenarios by using the proposed optimization method, which is given in Section V. The main parameters adopted here are set as $\sigma^2 = - 64$ dB, $d_0 = 1500$ km, $L = 15$ km,  $R_{\rm{E}} = 6371 $ km, $U_1 = 8371$ km, $U_2 = 6531$ km, $P_{{\rm{S}}}^{\max} = 50$ W, and $P_{{\rm{R}}}^{\max} = 10$ W. In the following two figures, for simplification purpose, we adopt "NI" and "IS" to represent non-interference and interference scenarios, respectively.

{\color{black}{We first evaluate the system performance for the proposed algorithm in terms of convergence. Fig. 13 depicts the evolution of energy efficiency for different permutations of $L$ and $H$. We can observe that the proposed Algorithm \ref{alg:1} achieves convergence within five iterations. } }

As observed in Figs. 14 and 15, energy efficiency gets worse while the height of the aerial relay, $H_1$, increases in both non-interference and interference scenarios. Because a large $H_1$ will result in large path-loss over R-D link, which leads to increased energy consumption for the information transmissions over R-D link. Moreover, it is obvious that the energy efficiency in the non-interference scenario is much higher than that in the interference scenario. This finding can be easily understood by the fact that more energy consumption is needed in the interference scenario to overcome the negative impact of the interfering signal on the information transmission over R-D link.

In Fig. 14, one can see that the energy efficiency with a small $L$ outperforms that with a large $L$ in both non-interference and interference scenarios. This observation can also be explained by the reason provided in the previous paragraph. Namely, a large $L$ will also incur large path-loss over R-D link, and more energy is required to realize the information delivery over R-D link.

From Fig. 15, we can easily find that the energy efficiency with a small $d_0$ outperforms that with a large $d_0$ in both considered scenarios. Because a large $d_0$ means a large transmission distance over S-R link, resulting in large path-loss over S-R link. Furthermore, it is seen from Fig. 15 that the energy efficiency for various $d_0 $ in the interference scenario is similar. Especially, when the height of R is too large (e.g., $H_1 > 20$ km), the differences among the energy efficiency for various $d_0$ are on the order of $10^7$ bits/Joule. {\color{black}{This observation can be explained the fact that the path loss over S-R link will dominate the main power consumption over S-R-D link when $d_0 $ is sufficiently large, e.g., the value of $d_0 $ ranges from 1600 km to 1800 km considered in Fig. 15, which is quite larger than the one of $H_1$ ranging from 5 km to 30 km.}}

{\color{black}{Fig. 16 indicates that the energy efficiency of the system increases with $T$ increasing in every considered scenario. With the time $T$ increasing, both satellite and relay can adopt a lower transmit power to achieve a better energy efficiency performance. For fixed $T$, the energy efficiency for non-interference scenario is one order higher than that for interference scenario.}}

It can be observed from Fig. 17 that the maximum power allowance (i.e., $P_S^{\max}$) has a negligible impact on the energy efficiency of the target system. {\color{black}{This observation can be explained as follows: the long-distance transmission over S-R link, which is normally on the order of 1000 km, leads to negligible improvements on the path-loss of S-R link, e.g., the path-loss of S-R link ranges from 3.5 $\times 10^{-11}$ to 6 $\times 10^{-11}$ when the maximum power allowance at S increases from 35 W to 60 W and path-loss factor is 2, and further exhibits no obvious influence on the transmission capacity of S-R link and the energy efficiency of the whole system.}} With our practical simulation setup, we find that the solution to the optimization problem always works in a reasonable region of the power allowance.

\section{Conclusion}
In this paper, we first have studied the outage performance of a CSATC system with DF relay scheme and derived the approximated analytical expressions for the coverage probability over R-D link and the e2e OP in both non-interference and interference scenarios, while considering the randomness of the positions of the satellite and the terrestrial receivers. Next, the e2e energy efficiency of the considered system has been optimized by employing the proposed optimization model.

Observing from the numerical results, some remarks can be achieved as follows:

1) The radius of the coverage space of the aerial relay shows a negative impact on the coverage/outage/energy efficiency performance in both non-interference and interference scenarios.

2) The radius of the distribution area of the interfering node exhibits a positive influence on the coverage/outage/energy efficiency performance.

3) The height of the aerial relay shows a clear negative effect on the coverage performance over R-D link and a weak impact on the outage performance over S-R link.

4) Adjusting the height of aerial relay cannot always improve the e2e outage/energy efficiency performance of the considered system, which has a performance limitation arisen from the transmission distance from the satellite to the aerial relay. Because the transmission distance over S-R link is quite larger than that over R-D link.

5) The size of the distribution space of the satellite shows an obvious negative effect on the outage performance over S-R link.

6) Increasing the transmit power at the satellite is not an efficient way to improve the performance of the whole system, due to the huge path-loss arisen from the long-distance transmission over S-R link normally on the order of thousands of kilometers.

\section*{Appendix: Proof of Lemma 2}
Let $t = \cos \theta $ ($ - 1 \le t \le 1$), we have $\theta  = arc\cos y$. Then, the PDF of $\cos \theta $ can be written as
\begin{align}
{f_{\cos \theta }}\left( t \right)& = \frac{2}{{2\pi }}\left| { - \frac{1}{{\sqrt {1 - {t^2}} }}} \right|\notag\\
 &= \frac{1}{{\pi \sqrt {1 - {t^2}} }}.
\end{align}

Therefore, considering the randomness of the position of ${\cos{\theta}}$ and using Chebyshev-Gauss quadrature in the first case, which is presented as $\int_{ - 1}^1 {\frac{{f\left( x \right)}}{{\sqrt {1 - {x^2}} }}dx}  \approx \sum\limits_{i = 1}^{{W_T}} {{w_i}f\left( {{x_i}} \right)} $ with ${x_i} = \cos \left( {\frac{{2i - 1}}{{2T}}\pi } \right)$ and the weight ${w_i} = \frac{\pi }{T}$, the CDF of ${\gamma _{{\rm{RD}}}}$ can be rewritten as
\begin{align}
{F_{{\gamma _{{\rm{RD}}}}}}\left( x \right) &= {\Gamma _1}\int \limits_{ - 1}^1 {{{\left( {{\Gamma _2} + \frac{{\Gamma _3}}{x}\frac{{r_i^2 + r_I^2 - 2{r_i}{r_I}t}}{{r_i^2 + {H_1}^2}}} \right)}^{ - 1}}} \notag\\
& \times {}_1{F_1}\left( {1;1;{\Gamma _4}{{\left( {{\Gamma _2} + \frac{{\Gamma _3}}{x}\frac{{r_i^2 + r_I^2 - 2{r_i}{r_I}t}}{{r_i^2 + {H_1}^2}}} \right)}^{ - 1}}} \right)\notag\\
& \times{f_{\cos \theta }}\left( t \right)dt\notag\\
& = \frac{{{\Gamma _1}}}{{\pi }}\sum\limits_{g = 1}^G {{\vartheta _g}} {\left( {{\Gamma _2} + \frac{{\Gamma _3}}{x}\frac{{r_i^2 + r_I^2 - 2{r_i}{r_I}{\varsigma _g}}}{{r_i^2 + {H_1}^2}}} \right)^{ - 1}}\notag\\
& \times {}_1{F_1}\left( {1;1;{\Gamma _4}{{\left( {{\Gamma _2} + \frac{{\Gamma _3}}{x}\frac{{r_i^2 + r_I^2 - 2{r_i}{r_I}{\varsigma _g}}}{{r_i^2 + {H_1}^2}}} \right)}^{ - 1}}} \right),
\end{align}
where ${\varsigma _g} = \cos \left( {\frac{{2g - 1}}{{2G}}\pi } \right)$ and ${\vartheta _g} = \frac{\pi }{G}$.

As shown in Fig. 2, ${\rm{D}}_i$ is uniformly distributed in the circle with original O and radius $L$, the PDF of $r_i$ can be given as
\begin{align}
 {f_{{r_i}}}\left( x \right)=\left\{ {\begin{array}{*{20}{l}}
{\frac{{2x}}{{{L^2}}},}&{{\rm{if ~ }}0 \le x \le L;}\\
{0,}&{{\rm{else}}}
\end{array}} \right.   .
\end{align}

Similarly, the PDF of $r_I$ can be written as
\begin{align}
   {f_{{r_I}}}\left( x \right)=\left\{ {\begin{array}{*{20}{l}}
{\frac{{2x}}{{{T^2}}},}&{{\rm{if  ~}}0 \le x \le T;}\\
{0,}&{{\rm{else}}}
\end{array}} \right. .
\end{align}

Similarly, taking the randomness of the position of I into account and applying Chebyshev-Gauss quadrature in the first case, the CDF of ${\gamma _{{\rm{RD}}}}$ can be presented as \eqref{CGI}, shown on the top of this page, where ${\eta _1} = \frac{T}{2}$, ${\zeta _j} = \cos \left( {\frac{{2j - 1}}{{2H}}\pi } \right)$, and ${\nu _j} = \frac{\pi }{H}$.
\begin{figure*}[ht]
\begin{align}\label{CGI}
{F_{{\gamma _{{\rm{RD}}}}}}\left( x \right)&  = \frac{{{2\Gamma _1}}}{{\pi {T^2}}}\sum\limits_{g = 1}^G {{\vartheta _g}\int_0^T {z{{\left( {{\Gamma _2} + \frac{{\Gamma _3}}{x}\frac{{r_i^2 + {z^2} - 2{r_i}z{\varsigma _g}}}{{r_i^2 + {H_1}^2}}} \right)}^{ - 1}}} }   {}_1{F_1}\left( {1;1;{\Gamma _4}{{\left( {{\Gamma _2} + \frac{{\Gamma _3}}{x}\frac{{r_i^2 + {z^2} - 2{r_i}z{\varsigma _g}}}{{r_i^2 + {H_1}^2}}} \right)}^{ - 1}}} \right)dz\notag\\
& = \frac{{{\Gamma _1}}}{{2\pi }}\sum\limits_{g = 1}^G {{\vartheta _g}\sum\limits_{j = 1}^H {{\zeta _j}\sqrt {1 - \nu _j^2} } } \left( {{\zeta _j} + 1} \right){\left( {{\Gamma _2} + \frac{{\Gamma _3}}{x}\frac{{r_i^2 + {\eta _1}^2{{\left( {{\zeta _j} + 1} \right)}^2} - 2{\eta _1}{r_i}\left( {{\zeta _j} + 1} \right){\varsigma _g}}}{{r_i^2 + {H_1}^2}}} \right)^{ - 1}}\notag\\
& ~~~~\times {}_1{F_1}\left( {1;1;{\Gamma _4}{{\left( {{\Gamma _2} + \frac{{\Gamma _3}}{x}\frac{{r_i^2 + {\eta _1}^2{{\left( {{\zeta _j} + 1} \right)}^2} - 2{\eta _1}{r_i}\left( {{\zeta _j} + 1} \right){\varsigma _g}}}{{r_i^2 + {H_1}^2}}} \right)}^{ - 1}}} \right)
\end{align}
\rule{18cm}{0.01cm}
\end{figure*}

Again, considering the randomness of the position of ${\rm{D}}_i$ and using Chebyshev-Gauss quadrature in the first case, the CDF of ${\gamma _{{\rm{RD}}}}$ can be further derived as \eqref{CGDi}, shown on the top of next page, where ${\eta _2} = \frac{L}{2}$, ${\kappa _i} = \cos \left( {\frac{{2i - 1}}{{2H}}\pi } \right)$, ${\iota _i} = \frac{\pi }{J}$, and ${\Theta _{g,j,k}}\left(x  \right) = {\left( {{\Gamma _2} + \frac{{\Gamma _3}}{x}\frac{{{\eta _2}^2{{\left( {{\kappa _k} + 1} \right)}^2} + {\eta _1}^2{{\left( {{\zeta _j} + 1} \right)}^2} - 2{\eta _1}{\eta _2}{\varsigma _g}\left( {{\zeta _j} + 1} \right)\left( {{\kappa _k} + 1} \right)}}{{{\eta _2}^2{{\left( {{\kappa _k} + 1} \right)}^2} + {H_1}^2}}} \right)^{ - 1}}$.
\begin{figure*}
\begin{align}\label{CGDi}
{F_{{\gamma _{{\rm{RD}}}}}}\left( x \right) &= \frac{{{\Gamma _1}}}{{\pi L^2}}\sum\limits_{g = 1}^G {{\vartheta _g} \sum\limits_{j = 1}^H {{\zeta _j}\sqrt {1 - \nu _j^2} } } \left( {{\zeta _j} + 1} \right)\int_0^L {{{\left( {{\Gamma _2} + \frac{{\Gamma _3}}{x}\frac{{{z^2} + {\eta _1}^2{{\left( {{\zeta _j} + 1} \right)}^2} - 2{\eta _1}{\varsigma _g}\left( {{\zeta _j} + 1} \right)z}}{{{z^2} + {H_1}^2}}} \right)}^{ - 1}}} \notag\\
&~~~~ \times {}_1{F_1}\left( {1;1;{\Gamma _4}{{\left( {{\Gamma _2} + \frac{{\Gamma _3}}{x}\frac{{{z^2} + {\eta _1}^2{{\left( {{\zeta _j} + 1} \right)}^2} - 2{\eta _1}{\varsigma _g}\left( {{\zeta _j} + 1} \right)z}}{{{z^2} + {H_1}^2}}} \right)}^{ - 1}}} \right){f_{{r_i}}}\left( z \right)dz\notag\\
& = \frac{{{\Gamma _1}}}{{4\pi}}\sum\limits_{g = 1}^G {{\vartheta _g} \sum\limits_{j = 1}^H {{\zeta _j}\sqrt {1 - \nu _j^2} } } \left( {{\zeta _j} + 1} \right)\int_{ - 1}^1 {\left( {y + 1} \right)} \notag\\
&~~~~\times {\left( {{\Gamma _2} + \frac{{\Gamma _3}}{x}\frac{{{\eta _2}^2{{\left( {y + 1} \right)}^2} + {\eta _1}^2{{\left( {{\zeta _j} + 1} \right)}^2} - 2{\eta _1}{\eta _2}{\varsigma _g}\left( {{\zeta _j} + 1} \right)\left( {y + 1} \right)}}{{{\eta _2}^2{{\left( {y + 1} \right)}^2} + {H_1}^2}}} \right)^{ - 1}}\notag\\
&~~~~\times {}_1{F_1}\left( {1;1;{\Gamma _4}{{\left( {{\Gamma _2} + \frac{{\Gamma _3}}{x}\frac{{{\eta _2}^2{{\left( {y + 1} \right)}^2} + {\eta _1}^2{{\left( {{\zeta _j} + 1} \right)}^2} - 2{\eta _1}{\eta _2}{\varsigma _g}\left( {{\zeta _j} + 1} \right)\left( {y + 1} \right)}}{{{\eta _2}^2{{\left( {y + 1} \right)}^2} + {H_1}^2}}} \right)}^{ - 1}}} \right)dy\notag\\
&=\frac{{{\Gamma _1}}}{{4\pi}}\sum\limits_{g = 1}^G {{\vartheta _g} \sum\limits_{j = 1}^H {{\zeta _j}\sqrt {1 - \nu _j^2} } } \left( {{\zeta _j} + 1} \right) \sum\limits_{k = 1}^J {{\iota _k}\sqrt {1 - \kappa _k^2} } \left( {{\kappa _k} + 1} \right){\Theta _{g,j,k}}\left(x  \right) {}_1{F_1}\left( {1;1;{\Gamma _4}{\Theta _{g,j,k}}\left(x  \right)} \right)
\end{align}
\rule{18cm}{0.01cm}
\end{figure*}

To facilitate the presentation, we adopt $\sum\limits_g {\sum\limits_j {\sum\limits_k {} } } $ to denote $\sum\limits_{g = 1}^G {{\vartheta _g}\sum\limits_{j = 1}^H {{\zeta _j}\sqrt {1 - \nu _j^2} } } \left( {{\zeta _j} + 1} \right)\sum\limits_{k = 1}^J {{\iota _k}\sqrt {1 - \kappa _k^2} } \left( {{\kappa _k} + 1} \right)$. Then, the CDF of ${\gamma _{{\rm{RD}}}}$ can be finally presented as \eqref{lemma2}.

So, the proof of Lemma 2 is finished.

\section{Appendix II: Proof of Theorem 2}
The volume of target space shown in Fig. 3, namely, the volume of the differece space between the spherical cones with radius $U_1$ and $U_2$ can be expressed as
\begin{align}
    V = \frac{{2\pi }}{3}\left( {1 - \cos \frac{\Psi }{2}} \right)  \left( {{U_1}^3 - {U_2}^3} \right).
\end{align}

Thus, employing the method adopted in Appendix A of \cite{PanTGCN19}, and considering that S is uniformly distributed, the CDF of $ \theta $ can be given as
\begin{align}
{F_\theta }\left( x \right) &= \int\limits_{{U_2}}^{{U_1}} {\int\limits_0^{2\pi } {\int\limits_0^x {\frac{1}{V}{d^2}\sin \theta d\theta d\varphi d\left( d \right)} } } \notag \\
{\rm{         }}& = \frac{1}{V}\frac{{2\pi }}{3}\left( {{U_1}^3 - {U_2}^3} \right)\left( {1 - \cos x} \right)\notag\\
{\rm{         }} &= \left\{ {\begin{array}{*{20}{l}}
{\frac{{1 - \cos x}}{{1 - \cos \frac{\Psi }{2}}};}&{{\rm{if~ }}0 \le x \le \frac{\Psi }{2}}\\
{0;}&{{\rm{else}}}
\end{array}} \right..
\end{align}

So, it is easy to have the PDF of $\theta$ as
\begin{align}
{f_\theta }\left( x \right) = \frac{{\sin x}}{{1 - \cos \frac{\Psi }{2}}}.
\end{align}

Similarly, employing the method adopted in Appendix A of \cite{PanTGCN19}, and considering that S is uniformly distributed, the PDF of $r_{\rm{S}}$ can be written as
\begin{align}
{f_{{r_{\rm{S}}}}}\left( x \right) = \left\{ {\begin{array}{*{20}{l}}
{\frac{{3{x^2}}}{{U_1^3 - U_2^3}}};&{{\rm{if  }}~{U_2} \le x \le {U_1}}\\
0;&{{\rm{else}}}
\end{array}} \right..
\end{align}

Thus, the joint PDF of $r_{\rm{S}}$ and $\theta$ can be presented as
\begin{align}
{f_{{r_{\rm{S}}},\theta }}\left( {x,y} \right) = \frac{{3{x^2}}}{{U_1^3 - U_2^3}}  \frac{{\sin y}}{{1 - \cos \frac{\Psi }{2}}},
\end{align}
where $0 \le \theta \le \Psi/2$ and ${{U_2} \le x \le {U_1}}$.

Since ${d_0} = \sqrt {r_{\rm{S}}^2 + H_{\rm{R}}^2 - 2{r_{\rm{S}}}{H_{\rm{R}}}\cos \theta } $, one can have
\begin{align}
\left| {\frac{{\partial \left( {d_0^2,{r_{\rm{S}}}} \right)}}{{\partial \left( {{r_{\rm{S}}},\theta } \right)}}} \right| &= \left| {\begin{array}{*{20}{c}}
{2{r_{\rm{S}}} - 2{H_{\rm{R}}}\cos \theta }&{2{r_{\rm{S}}}{H_{\rm{R}}}\sin \theta }\\
1&0
\end{array}} \right|\notag\\
{\rm{               }} &= 2{r_{\rm{S}}}{H_{\rm{R}}} {\sin \theta }.
\end{align}

Then, the joint PDF of $d_0^2$ and ${r_{\rm{S}}}$ can be written as
\begin{align}
{f_{d_0^2,{r_{\rm{S}}}}}\left( {x,y} \right) &= \frac{{{f_{{r_{\rm{S}}},\theta }}\left( {x,y} \right)}}{{\left| {\frac{{\partial \left( {d_0^2,{r_{\rm{S}}}} \right)}}{{\partial \left( {{r_{\rm{S}}},\theta } \right)}}} \right|}} \notag\\
{\rm{                 }} &= \frac{1}{{2{H_{\rm{R}}}}} \frac{1}{{1 - \cos \frac{\Psi }{2}}}\frac{{3y}}{{U_1^3 - U_2^3}},
\end{align}
where ${{U_2} }\le y \le {U_1}$ and $\cos \frac{\Psi }{2} \le \frac{{{y^2} + H_{\rm{R}}^2 - x}}{{2y{H_{\rm{R}}}}} \le 1$.

So, we can calculate the PDF of $d_0^2$ as
\begin{align}
{f_{d_0^2}}\left( x \right) &= \frac{3}{{4{H_{\rm{R}}}}}  \frac{1}{{1 - \cos \frac{\Psi }{2}}}\frac{1}{{U_1^3 - U_2^3}}\left[ \omega^2 (x) - \rho^2 (x) \right]
\notag\\
&=\tau  \left[ \omega ^2(x) - \rho^2 (x) \right] ,
\end{align}
where $\omega (x) = \min \left\{ {{U_1},{H_{\rm{R}}} + \sqrt x } \right\}$, $\rho (x) = \max \left\{ {{U_2},{H_{\rm{R}}}\cos \frac{\psi }{2} + \sqrt {x - {H_{\rm{R}}}^2{{\sin }^2}\frac{\psi }{2}} } \right\}$, $ \tau = \frac{3}{{4{H_{\rm{R}}}}} \frac{1}{{1 - \cos \frac{\Psi }{2}}}\frac{1}{{U_1^3 - U_2^3}}$ and ${x - {H_{\rm{R}}}^2{{\cos }^2}\frac{\Psi }{2} \ge 0}$.

With this result, the proof of Theorem 2 is completed.


\begin{thebibliography}{1}

\bibitem{Erdelj}
M. Erdelj, E. Natalizio, K. R. Chowdhury, and I. F. Akyildiz, ``Help from the sky: Leveraging UAVs for disaster management," \emph{IEEE Pervasive Comput.}, vol. 16, no. 1, pp. 24-32, Jan. 2017.



\bibitem{YRuan}
Y. Ruan, Y. Li, C. Wang, and R. Zhang, ``Energy efficient adaptive transmissions in integrated satellite-terrestrial networks with SER constraints," \emph{IEEE Trans. Wireless Commun.}, vol. 17, no. 1, pp. 210-222, Jan. 2018.

\bibitem{AMK}
Arti M. K., ``A novel beamforming and combining scheme for two-way AF satellite systems," \emph{IEEE Trans. Veh. Technol.}, vol. 66, no. 2, pp. 1248-1256, Feb. 2017.

\bibitem{AMK3}
Arti M. K. and S. K. Jindal, ``OSTBC transmission in shadowed-Rician land mobile satellite links," \emph{IEEE Trans. Veh. Technol.}, vol. 65, no. 7, pp. 5771-5777, July 2016.

\bibitem{LYang}
L. Yang and M. O. Hasna, ``Performance analysis of amplify-and-forward hybrid satellite-terrestrial networks with cochannel interference," \emph{IEEE Trans. Commun.}, vol. 63, no. 12, pp. 5052-5061, Dec. 2015.

\bibitem{KAn}
K. An et al., ``Performance analysis of multi-antenna hybrid satellite-terrestrial relay networks in the presence of interference," \emph{IEEE Trans. Commun.}, vol. 63, no. 11, pp. 4390-4404, Nov. 2015.

\bibitem{KAn2}
K. An, M. Lin, J. Ouyang, et al., ``Symbol error analysis of hybrid satellite¨Cterrestrial cooperative networks with cochannel interference," \emph{IEEE Commun. Lett.}, vol. 18, no. 11, pp. 1947-1950, Nov. 2014.

\bibitem{Bhatnagar}
M. R. Bhatnagar and Arti M.K., ``Performance analysis of hybrid satellite-terrestrial FSO cooperative system," \emph{IEEE Photon. Technol. Lett.}, vol. 25, no. 22, pp. 2197-2200, Nov. 2013.

\bibitem{Bhatnagar2}
M. R. Bhatnagar and Arti M.K., ``Performance analysis of AF based hybrid satellite-terrestrial cooperative network over generalized fading channels," \emph{IEEE Commun. Lett.}, vol. 17, no. 10, pp. 1912-1915, Oct. 2013.

\bibitem{Sreng}
S. Sreng, B. Escrig, and M. Boucheret, ``Exact symbol error probability of hybrid/integrated satellite-terrestrial cooperative network," \emph{IEEE Trans. Wireless Commun.}, vol. 12, no. 3, pp. 1310-1319, Mar. 2013.










\bibitem{XYan}
X. Yan, H. Xiao, K. An, G. Zheng and W. Tao, ``Hybrid satellite terrestrial relay networks with cooperative non-orthogonal multiple access," \emph{IEEE Commun. Lett.}, vol. 22, no. 5, pp. 978-981, May 2018.

\bibitem{Sharma}
P. K. Sharma, P. K. Upadhyay, D. B. da Costa, et al., ``Performance analysis of overlay spectrum sharing in hybrid satellite-terrestrial systems with secondary network selection," \emph{IEEE Trans. Wireless Commun.}, vol. 16, no. 10, pp. 6586-6601, Oct. 2017.

\bibitem{Cocco}
G. Cocco, N. Alagha and C. Ibars, ``Cooperative coverage extension in vehicular land mobile satellite networks," \emph{IEEE Trans. Veh. Technol.}, vol. 65, no. 8, pp. 5995-6009, Aug. 2016.










\bibitem{Bankey}
V. Bankey and P. K. Upadhyay, ``Physical layer security of multiuser multirelay hybrid satellite-terrestrial relay networks," \emph{IEEE Trans. Veh. Technol.}, vol. 68, no. 3, pp. 2488-2501, Mar. 2019.

\bibitem{YAi}
Y. Ai, A. Mathur, M. Cheffena, M. R. Bhatnagar and H. Lei, ``Physical layer security of hybrid satellite-FSO cooperative systems," \emph{IEEE Photon. J.}, vol. 11, no. 1, pp. 1-14, Feb. 2019.

\bibitem{JDu}
J. Du, C. Jiang, H. Zhang, et al., ``Secure satellite-terrestrial transmission over incumbent terrestrial networks via cooperative beamforming," \emph{IEEE J. Sel. Areas Commun.}, vol. 36, no. 7, pp. 1367-1382, July 2018.

\bibitem{BLi}
B. Li, Z. Fei, Z. Chu, et al., ``Robust chance-constrained secure transmission for cognitive satellite¨Cterrestrial networks," \emph{IEEE Trans. Veh. Technol.}, vol. 67, no. 5, pp. 4208-4219, May 2018.

\bibitem{BLi2}
B. Li, Z. Fei, X. Xu and Z. Chu, ``Resource allocations for secure cognitive satellite-terrestrial networks," \emph{IEEE Wireless Commun. Lett.}, vol. 7, no. 1, pp. 78-81, Feb. 2018.









\bibitem{PIMRC2016}
Y. Xu, Y. Wang, R. Sun and Y. Zhang, ``Joint relay selection and power allocation for maximum energy efficiency in hybrid satellite-aerial-terrestrial systems," in \emph{Proc. PIMRC 2016}, Valencia, 2016, pp. 1-6.






\bibitem{Zhangyq}
Y. Zhang, J. Ye, G. Pan, and M. Alouini, ``Secrecy outage analysis for satellite-terrestrial downlink transmissions," to appear in \emph{IEEE Wireless Commun. Lett.}, doi: 10.1109/LWC.2020.2999555.



\bibitem{Abdi}
A. Abdi, W. Lau, M.-S. Alouini, and M. Kaveh, ``A new simple model for land mobile satellite channels: First and second order statistics" \emph{IEEE Trans. Wireless Commun.}, vol. 2, no. 3, pp. 519-528, May 2003.






\bibitem{GPanCL2017}
G. Pan, J. Ye, Z. Ding, ``On secure VLC systems with spatially random terminals," \emph{IEEE Commun. Lett.}, vol. 21, no. 3, pp. 492-495, Mar. 2017.

\bibitem{YetWCL}
J. Ye, C. Zhang, H. Lei, G. Pan, Z. Ding, ``Secure UAV-to-UAV systems with spatially random UAVs," \emph{IEEE Wireless Commun. Lett.},  vol. 8, no. 2, pp. 564-567, Apr. 2019.











\bibitem{Gradshteyn}
I.S. Gradshteyn and I.M. Ryzhik, \emph{Table of Integrals, Series and Products, 7 Ed}. San Diego: Academic Press, 2007.

\bibitem{Simon}
M. K. Simon and M.-S. Alouini, \emph{Digital Communication Over Fading Channels, 2nd ed.}, New York, NY, USA: Wiley, 2005.

\bibitem{Bocus}
M. Z. Bocus, C. P. Dettmann, and J. P. Coon, ``An approximation of the first order Marcum $Q$-function with application to network connectivity analysis, \emph{IEEE Commun. Lett.}, vol. 17, no. 3, pp. 499-502, Mar. 2013.


\bibitem{Prudnikov2}
A. P. Prudnikov, Y. A. Brychkov, and O. I. Marichev, \emph{Integrals and Series: Special Functions, vol. 2, 3rd ed.}, New York: Gordon \& Breach Sci. Publ., 1992.



























\bibitem{PanTcom2017}
G. Pan, H. Lei, Y. Yuan, Z. Ding, ``Performance analysis and optimization for SWIPT wireless sensor networks," \emph{IEEE Trans. Commun.}, vol. 65, no. 5, pp. 2291-2302, May 2017.

\bibitem{Boyd}
S. Boyd and L. Vandenberghe, \emph{Convex Optimization}. New  York, NY, USA: Cambridge University Press, 2004


\bibitem{PanTGCN19}
G. Pan, H. Lei, Z. Ding and Q. Ni, ``3-D hybrid VLC-RF indoor IoT systems with light energy harvesting," \emph{IEEE Trans. Green Commun. Netw.}, vol. 3, no. 3, pp. 853-865, Sept. 2019.

\end{thebibliography}
\end{document}